\documentclass[structabstract
]{aa}  
\usepackage{supertabular}
\usepackage{graphicx,natbib}
\usepackage{txfonts}
\usepackage{dcolumn}
\usepackage{lscape}

\newcommand{\um}{\,\mu\rm{m}}

\newcommand{\Spitzer}{{\it Spitzer}}                    
\newcommand{\Herschell}{{\it Herschel}\ }
\newcommand{\Herschel}{{\it Herschel}}

\newcommand{\dma}[1]{_{\mathrm{#1}}}

\begin{document}
   \title{A peculiar class of debris disks from Herschel/DUNES\thanks{\Herschell is an ESA space observatory with science instruments provided by European-led Principal Investigator consortia and with important participation from NASA.}}
   \subtitle{A steep fall off in the far infrared}
   \author{S. Ertel\inst{1,2}
        \and
          S. Wolf\inst{2}
        \and
          J.~P. Marshall\inst{3}
        \and
          C. Eiroa\inst{3}
        \and
          J.-C. Augereau\inst{1}
        \and
          A.~V. Krivov\inst{4}
        \and
          T. L\"ohne\inst{4}
        \and
          O. Absil\inst{5}
        \and
          D. Ardila\inst{6}
        \and
          M. Ar\'evalo\inst{7}
        \and
          A. Bayo\inst{8}
        \and
          G. Bryden\inst{9}
        \and
          C. del Burgo\inst{10}
        \and
          J. Greaves\inst{11}
        \and
          G. Kennedy\inst{12}
        \and
          J. Lebreton\inst{1}
        \and  
          R. Liseau\inst{13}
        \and
          J. Maldonado\inst{3}
        \and
          B. Montesinos\inst{8}
        \and
          A. Mora\inst{14}
        \and
          G.~L. Pilbratt\inst{15}
        \and
          J. Sanz-Forcada\inst{8}
        \and
          K. Stapelfeldt\inst{16}
        \and
          G.~J. White\inst{17,18}
   }
   \institute{UJF-Grenoble 1 / CNRS-INSU, Institut de Plan\'etologie et d'Astrophysique de Grenoble (IPAG) UMR 5274, Grenoble, F-38041, France \\
          \email{steve.ertel@obs.ujf-grenoble.fr} 
        \and
          Institut f\"ur Theoretische Physik und Astrophysik, Christian-Albrechts-Universit\"at zu Kiel, Leibnizstra{\ss}e 15, 24098 Kiel, Germany
        \and
          Dpt. F\'\i sica Te\'orica, Facultad de Ciencias, Universidad Aut\'onoma de Madrid, Cantoblanco, 28049 Madrid, Spain
        \and
          Astrophysikalisches Institut und Universit{\"a}tssternwarte, Friedrich-Schiller-Universit{\"a}t, Schillerg{\"a}{\ss}chen 2-3, 07745 Jena, Germany
        \and
          Institut d'Astrophysique et de G{\'e}ophysique, Universit{\'e} de Li{\`e}ge, 17 All{\'e}e du Six Ao{\^u}t, 4000 Sart Tilman, Belgium
        \and
          NASA Herschel Science Center, California Institute of Technology, 1200 E. California Blvd., Pasadena, CA 91125, USA
        \and
          Department of Astrophysics, Centre for Astrobiology (CAB, CSIC-INTA), ESAC Campus, P.O. Box 78, 28691 Villanueva de la Ca\~nada, Madrid, Spain.
        \and
          European Space Observatory, Alonso de Cordova 3107, Vitacura Casilla 19001, Santiago 19, Chile
        \and
          Jet Propulsion Laboratory, California Institute of Technology, Pasadena, CA 91109, USA
        \and
          UNINOVA-CA3, Campus da Caparica, Quinta da Torre, Monte de Caparica, 2825-149 Caparica, Portugal
        \and
          School of Physics and Astronomy, St Andrews University, North Haugh, St Andrews, Fife KY16 9SS
        \and
          Institute of Astronomy, University of Cambridge, Madingley Road, Cambridge CB3 0HA
        \and
          Onsala Space Observatory, Chalmers University of Technology, 439 92 Onsala, Sweden
        \and
          ESA-ESAC Gaia SOC. P.O. Box 78, 28691 Villanueva de la Ca{\~n}ada, Madrid, Spain
        \and
          ESA Astrophysics \& Fundamental Physics Missions Division, ESTEC/SRE-SA,  Keplerlaan 1, 2201 AZ Noordwijk, The Netherlands
        \and
          Code 667, NASA Goddard Space Flight Center, Greenbelt MD 20771 USA
        \and
          Department of Physics and Astrophysics, Open University, Walton Hall, Milton Keynes MK7 6AA, UK
        \and
          Rutherford Appleton Laboratory, Chilton OX11 0QX, UK
   }
   \date{}

 
  \abstract
   {The existence of debris disks around old main sequence stars is usually explained by continuous replenishment of small dust grains through collisions from a reservoir of larger objects.}
   {We present photometric data of debris disks around HIP\,103389 (HD\,199260), HIP\,107350 (HN\,Peg, HD\,206860), and HIP\,114948 (HD\,219482), obtained in the context of our \Herschell Open Time Key Program DUNES (DUst around NEarby Stars).}
   {We used \Herschel/PACS to detect the thermal emission of the three debris disks with a $3\sigma$ sensitivity of a few mJy at $100\um$ and $160\um$. In addition, we obtained \Herschel/PACS photometric data at $70\um$ for HIP\,103389. These observations are complemented by a large variety of optical to far-infrared photometric data. Two different approaches are applied to reduce the \Herschell data to investigate the impact of data reduction on the photometry. We fit analytical models to the available spectral energy distribution (SED) data using the fitting method of simulated thermal annealing as well as a classical grid search method.}
   {The SEDs of the three disks potentially exhibit an unusually steep decrease at wavelengths \mbox{$\ge 70\,\mu\rm{m}$}. We investigate the significance of the peculiar shape of these SEDs and the impact on models of the disks provided it is real. Using grain compositions that have been applied successfully for modeling of many other debris disks, our modeling reveals that such a steep decrease of the SEDs in the long wavelength regime is inconsistent with a power-law exponent of the grain size distribution $-3.5$ expected from a standard equilibrium collisional cascade. In contrast, a steep grain size distribution or, alternatively an upper grain size in the range of few tens of micrometers are implied. This suggests that a very distinct range of grain sizes would dominate the thermal emission of such disks. However, we demonstrate that the understanding of the data of faint sources obtained with \Herschell is still incomplete and that the significance of our results depends on the version of the data reduction pipeline used.}
   {A new mechanism to produce the dust in the presented debris disks, deviations from the conditions required for a standard equilibrium collisional cascade (grain size exponent of $-3.5$), and/or significantly different dust properties would be necessary to explain the potentially steep SED shape of the three debris disks presented.}


   \maketitle

\section{Introduction}
\label{intro}

Debris disks were first discovered via infrared excess emission associated with main sequence stars such as Vega detected by the {\it Infrared Astronomical Satellite} \citep[IRAS;][]{aum84}. The first spatially resolved image of a debris disk was that of $\beta$\,Pictoris \citep{smi84} in optical scattered light. In the last few years, the {\it Spitzer Space Telescope} has revealed that debris disks are common around main sequence stars \citep[e.g.,][]{tri08}. Several Key Programs on the {\it Herschel Space Observatory} \citep{pil10} are dedicated to the study of various aspects of the formation and evolution of planetary systems and their attendant circumstellar debris disks \citep[e.g.,][]{aug08}. Our \Herschell Open Time Key Program (OTKP) DUNES \citep[DUst around NEarby Stars;][]{eir10} aims to detect debris disks with fractional luminosities similar to the Edgeworth-Kuiper Belt level \citep[\mbox{$L_{\rm{d}}/L_{\star} = 10^{-7}$ to $10^{-6}$};][]{ste96, vit10} around a volume limited sample \mbox{($d \le 20\,\rm{pc}$)} of Sun-like stars (F, G, and K spectral type). Some additional sources at \mbox{$20\,{\rm pc} < d \le 25\,{\rm pc}$} are included because of their known excesses that have previously been detected with \Spitzer, or because they are known exoplanet host stars.

As the most readily detectable signposts of other planetary systems, debris disks help us to improve our understanding of the formation and evolution of them as well as of our own solar system \citep{mey07, wya08, kri10}. Studying the spectral energy distribution (SED) of the dust alone usually provides only weak, ambiguous constraints to their properties such as chemical composition, grain size, and spatial distribution \citep{wol03}. For example, the location of the inner disk radius of the dust distribution is strongly degenerate with the lower limit of the grain size distribution. However, the grain size distribution can be described by a power-law and the power-law exponent derived from pure SED fitting is usually well constrained and consistent with the analytical value of $-3.5$ derived by \citet{doh69} under several assumptions: (1) The particles are produced through a collisional cascade from infinitely large to extremely small grains. (2) The grains are not affected by any other influence such as stellar radiation and drag forces. (3) The strength of the particles, i.e., the energy per volume fraction necessary to disrupt them, is independent from the grain size. This is further referred to as a standard equilibrium collisional cascade.

In this paper, the potential \Herschel/DUNES discovery of an unusually steep decrease of the SEDs of three spatially unresolved debris disks around Sun-like stars, HIP\,103389 (HD\,199260), HIP\,107350 (HN\,Peg, HD\,206860), and HIP\,114948 (HD\,219482), is presented\footnote{Two of the sources, HIP\,103389 and HIP\,107350, are shared targets between the DUNES survey in the context of which the analysis is carried out and the DEBRIS survey \citep{mat10,phi10}.}. It is demonstrated on the example of the three targets that the analysis of data of faint point sources obtained with \Herschell depends very much on the photometric calibration and the exact determination of the uncertainties. In particular, the significance of our results depends on the version of the data reduction pipeline used. Provided it is real, the steep decrease occurs in the range of $70\um$ to $160\um$, inconsistent with a dust grain size distribution following a power-law derived from a standard equilibrium collisional cascade. The $100\um$ and $160\um$ data represent the first photometric measurements of these faint disks at wavelengths \mbox{$> 70\um$} and, thus, are the first ones to reveal the unusual shape of the SEDs. Results from detailed SED modeling including additional photometric data from the literature are presented as well.

Observations and data reduction are described in Sect.~\ref{obs+red} together with the basic observational results. A theoretical discussion on how steep the decrease of the SED of a debris disk toward longer wavelengths is expected to be and a characterization of unusually steep SED sources are given in Sect~\ref{theory}. A detailed description of our modeling of the systems can be found in Sect.~\ref{modeling} and results are discussed in Sect.~\ref{results}. Conclusions are given in Sect.~\ref{conc}.

\begin{figure}
\centering
\includegraphics[width=1\linewidth]{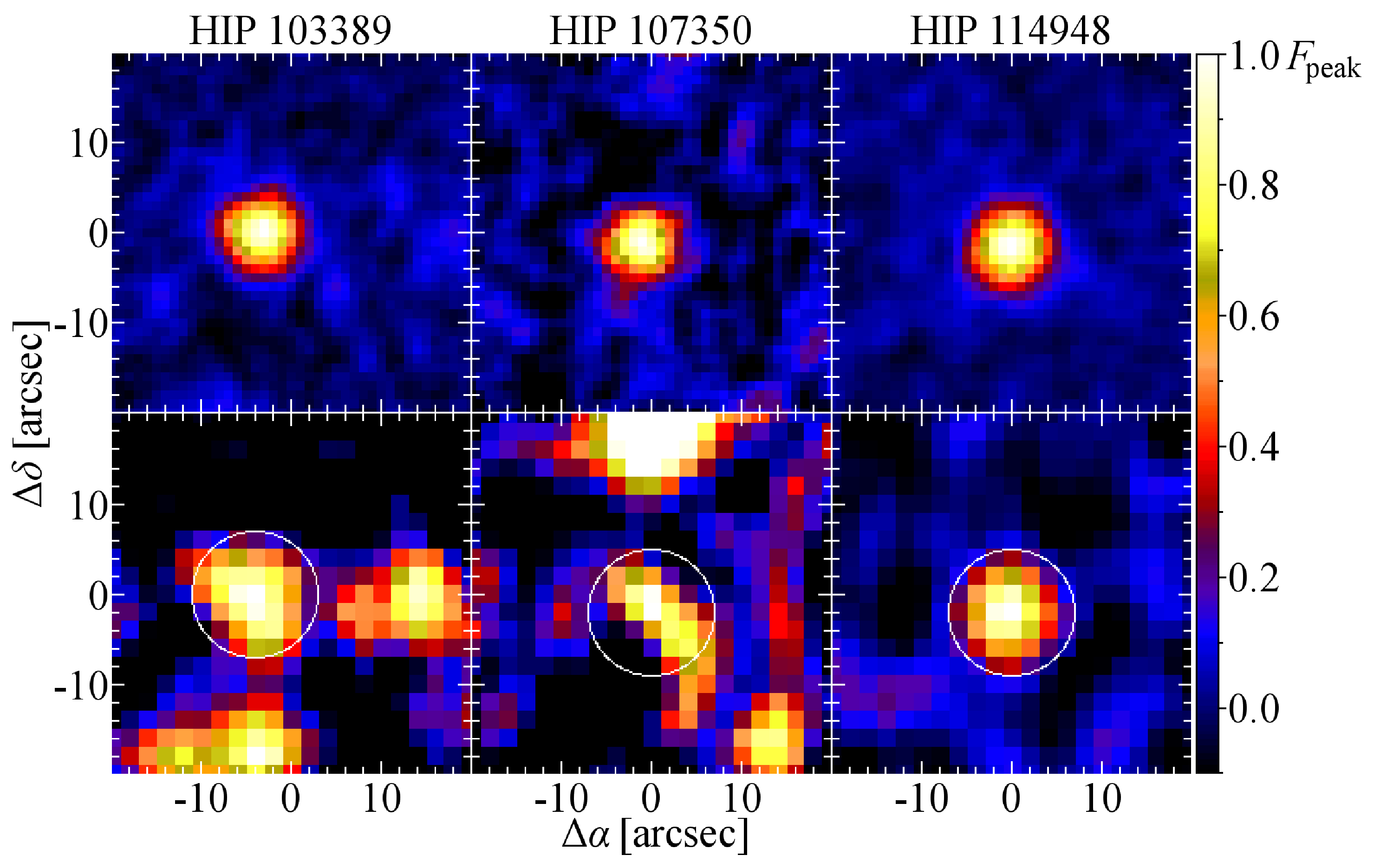}
\caption{Images of the three sources at $100\um$ (\emph{top}) and $160\um$ (\emph{bottom}) from Reduction~1 (Sect.~\ref{obs+red}). The images are displayed in a linear stretch from $-0.1\,F_{\rm peak}$ to $F_{\rm peak}$ of the source of interest. The peak flux is $0.38\,{\rm mJy/pixel}$, $0.23\,{\rm mJy/pixel}$, and $0.54\,{\rm mJy/pixel}$ at $100\um$ and $0.17\,{\rm mJy/pixel}$, $0.16\,{\rm mJy/pixel}$, and $0.45\,{\rm mJy/pixel}$ at $160\um$ for HIP\,103389, HIP\,107350, and HIP\,114948, respectively. The white circles in the $160\um$ images are centered on the brightest pixel of the source, respectively.}
\label{images_red}%
\end{figure}

\section{Observations \& data reduction}
\label{obs+red}

Two \Herschel/PACS mini-scan map observations of each target were taken with the 100/160 channel combination at array orientation angles of $70\degr$ and $110\degr$ providing scan and cross-scan coverage to assist in the removal of noise artifacts from the final composite mosaic. In addition, two scan map observations of HIP\,103389 were taken with the 70/160 channel combination with the same array orientations. Each scan map consists of 10 legs of $3'$ length, with a $4''$ separation between legs, at the medium slew speed ($20''$ per second). In this way, a region of \mbox{$\approx 1$} square arc minute around the source position was covered to uniform depth in the resulting mosaic. A summary of the observing log is presented in Table~\ref{obs_log}.

PACS data reduction was carried out in  version 4.2 ({\it Reduction~1}) and 7.2 ({\it Reduction~2}) of HIPE \citep{ott10} starting from the level 0 products using a modified version of the standard reduction script (provided within HIPE). Two different reductions have been carried out, because changes in the still developing pipeline of HIPE may significantly affect the results. The separate scans at the two position angles of each channel pair were mosaicked to produce a final image at each wavelength. Output scales for the final mosaics are $1''$ per pixel for the $70\um$ and $100\um$ images and $2''$ per pixel for the $160\um$ image (smaller than the native detector pixel sizes of $3\farcs2$ at $70\um$ and $100\um$ and $6\farcs4$ at  
$160\um$, respectively).

For the further analysis of the data, different approaches have been used for Reductions~1 and~2. These are described in the following:

\smallskip
\noindent{\bf Reduction~1:} A high-pass filter was used to remove large scale background variation from the images, with filter widths of 31 frames at $70\um$ and $100\um$ and 51 frames at $160\um$ (equivalent to $\sim 60''$ and $100''$, respectively). A central region of $30''$ radius in the images was masked from the high pass filter process to prevent the removal of any faint, extended structure near the source position. PACS fluxes and sky noise were measured by aperture photometry using a custom script based on the IDL APER routine, based on the DAOPHOT aperture photometry module and cross-checked with the internal HIPE aperture photometry routines. The aperture radius and sky annulus dimensions were $20''$ and $30'' - 40''$, respectively at all bands. Sky noise for each wavelength was calculated from the rms variance of the sky annulus pixel values multiplied by the beam size. Results were scaled by the appropriate aperture correction. Absolute uncertainties are 5\% at $70\um$ and $100\um$ and $10\%$ at $160\um$ using the calibration files provided along with this HIPE version.

The images of the three sources from Reduction~1 at $100\um$ and $160\um$ are shown in Fig.~\ref{images_red}.

\smallskip
\noindent{\bf Reduction~2:} The same filter width as in Reduction~1 has been used for the high-pass filtering, but sources were masked based on a threshold criterion, the value of which was calculated from the standard deviation of all non-zero pixels in the image. PACS fluxes and sky noise were measured by aperture photometry using both the internal HIPE aperture photometry routines and the MIDAS data analysis package to check for consistency. The source flux was measured within a circular aperture of $4''$, $5''$, and $8''$ radius at $70\um$, $100\um$, and $160\um$, respectively, and scaled by the appropriate aperture correction. Sky noise for each wavelength was calculated from the rms variance of ten sky apertures of the same size as the source aperture and randomly distributed across the uniformly covered part of the image. The resulting deviation of these apertures was calibrated by the aperture correction and correlated noise correction factors. Extensive testing has shown that this approach efficiently accounts for the correlated noise in the output images. In fact, this approach provides slightly conservative uncertainties. An extended discussion will be included in the survey overview paper (Eiroa et al., in prep.). Aperture correction factors have also been applied to the rms aperture flux dispersion. Photometric calibration uncertainties in this version of HIPE are $3\% - 5\%$\footnote{Release note PICC-ME-TN-037}.

\subsection{Observational results}
\label{obs_res}

\begin{table}
\caption{Summary of {\it Herschel}/PACS observations of the three objects}             
\label{obs_log}      
\centering          
\begin{tabular*}{1.0\columnwidth}{l@{\extracolsep{\fill}}cccr} 
\hline\hline
  Object    & Observation ID & OD  & $\lambda$ [$\um$] & $T$ [s]  \\
\hline
  HIP\,103389 & 1342193157/58  & 322 &  70/160           &  180   \\
  HIP\,103389 & 1342193159/60  & 322 & 100/160           & 1440   \\                             
  HIP\,107350 & 1342195779/80  & 355 & 100/160           & 1440   \\                                         
  HIP\,114948 & 1342196803/04  & 372 & 100/160           & 1440   \\
\hline                  
\end{tabular*}
\tablefoot{OD is the observing date that counts the elapsed time in days since launch on 14 May 2009, $\lambda$ gives the central wavelength of the two filters used simultaneously, and $T$ is the total on-source integration time.}
\end{table}

\begin{table*}
\caption{Observational results obtained in this work\tablefootmark{a}}
\label{herscheldata}
\begin{center}
\begin{tabular*}{0.8\linewidth}{l@{\extracolsep{\fill}}cccccc}
\hline\hline
 Source              & \multicolumn{2}{c}{HIP\,103389} & \multicolumn{2}{c}{HIP\,107350} & \multicolumn{2}{c}{HIP\,114948} \\
 \cline{2-3} \cline{4-5} \cline{6-7}
  & Reduction~1 & Reduction~2 & Reduction~1 & Reduction~2 & Reduction~1 & Reduction~2 \\
\hline
 $F_{\rm PACS\,70}$ [mJy] & $ 47.4 \pm 2.7 $  & $ 44.0 \pm 2.3 $ &             \dots & \dots &             \dots & \dots \\
 $F_{\rm MIPS\,70}$ [mJy] & \multicolumn{2}{c}{$ 46.6 \pm 3.8 $} & \multicolumn{2}{c}{$ 28.4 \pm 2.5 $} & \multicolumn{2}{c}{$ 68.7 \pm 3.0 $} \\
 $F_{\rm PACS\,100}$ [mJy] & $ 23.7 \pm 1.4 $  & $ 26.3 \pm 1.7 $ & $ 11.0 \pm 0.9 $ & $ 15.1 \pm 1.3 $ &  $ 42.5 \pm 2.2 $ & $ 40.8 \pm 1.6 $ \\
 $F_{\rm PACS\,160}$ [mJy] & $  5.0 \pm 1.3 $  & $  7.7 \pm 2.5 $ & $  4.4 \pm 1.5 $ & $  4.4 \pm 2.3 $ &  $ 12.7 \pm 1.9 $ & $ 13.3 \pm 2.2 $ \\
\hline
 $F_{\star, 70\um}$\tablefootmark{b} [mJy]  & \multicolumn{2}{c}{$13.8$} & \multicolumn{2}{c}{$13.4$} & \multicolumn{2}{c}{$15.0$} \\
 $F_{\star, 100\um}$\tablefootmark{b} [mJy] & \multicolumn{2}{c}{$6.8$} & \multicolumn{2}{c}{$6.6$} & \multicolumn{2}{c}{$7.3$} \\
 $F_{\star, 160\um}$\tablefootmark{b} [mJy] & \multicolumn{2}{c}{$2.7$} & \multicolumn{2}{c}{$2.6$} & \multicolumn{2}{c}{$2.9$} \\
\hline
 $\Delta_{100,70}$               & $ 1.94 \pm 0.32 $ & $ 1.44 \pm 0.33 $ & $ 2.66 \pm 0.45 $ & $ 1.77 \pm 0.49 $ & $ 1.35 \pm 0.26 $ & $ 1.46 \pm 0.23 $ \\
 $\Delta_{160,100}$              & $ 3.31 \pm 0.69 $ & $ 2.61 \pm 0.82 $ & $ 1.95 \pm 0.88 $ & $ 2.62 \pm 1.30 $ & $ 2.57 \pm 0.44 $ & $ 2.38 \pm 0.43 $ \\
 $\Delta_{160,70}$               & $ 2.72 \pm 0.39 $ & $ 2.11 \pm 0.46 $ & $ 2.25 \pm 0.53 $ & $ 2.26 \pm 0.74 $ & $ 2.04 \pm 0.22 $ & $ 1.99 \pm 0.25 $ \\
\hline
 ${\rm FWHM}_{\rm PACS\,100}$\tablefootmark{c}   & \multicolumn{2}{c}{$6\farcs8\times7\farcs1$} & \multicolumn{2}{c}{$7\farcs1\times6\farcs0$} & \multicolumn{2}{c}{$7\farcs2\times7\farcs2$} \\
\hline
 ${\rm RA}_{\rm PACS\,100}$      & \multicolumn{2}{c}{$20^{\rm h}56^{\rm m}47\fs47$} & \multicolumn{2}{c}{$21^{\rm h}44^{\rm m}31\fs27$} & \multicolumn{2}{c}{$23^{\rm h}16^{\rm m}57\fs46$} \\
 ${\rm DEC}_{\rm PACS\,100}$     & \multicolumn{2}{c}{$ -26^\circ17'45\farcs5 $} & \multicolumn{2}{c}{$ +14^\circ46'19\farcs4 $} & \multicolumn{2}{c}{$ -62^\circ00'05\farcs7 $} \\
 ${\rm RA}_{\rm optical}$        & \multicolumn{2}{c}{$20^{\rm h}56^{\rm m}47\fs33$} & \multicolumn{2}{c}{$21^{\rm h}44^{\rm m}31\fs33$} & \multicolumn{2}{c}{$23^{\rm h}16^{\rm m}57\fs69$} \\
 ${\rm DEC}_{\rm optical}$       & \multicolumn{2}{c}{$ -26^\circ17'47\farcs0 $} & \multicolumn{2}{c}{$ +14^\circ46'19\farcs0 $} & \multicolumn{2}{c}{$ -62^\circ00'04\farcs3 $} \\
\hline
\end{tabular*}
\end{center}
\tablefoot{Uncertainties of the flux measurements are total uncertainties including sky noise and calibration uncertainties. $\Delta_{\nu_1, \nu_2}$ is the spectral index measured from $\nu_1$ and $\nu_2$ (Sect~\ref{theory}), identified with the corresponding PACS bands.\\
\tablefoottext{a}{The \Spitzer/MIPS $70\um$ photometry is also listed, since these data complement the \Herschell observations and are important to illustrate the unusual behavior of the SEDs. These data are published by \citet[][HIP\,103389 and HIP\,114948]{bei06} and \citet[][HIP\,107350]{bry06} and have been re-reduced in the context of DUNES (Eiroa et al., in prep.).}
\tablefoottext{b}{Predicted stellar photosphere using a PHOENIX/GAIA synthetic stellar model (Eiroa et al., in prep.).}
\tablefoottext{c}{FWHM as measured from the PACS images at $100\um$ using a 2-D Gaussian fit.}
}
\end{table*}

The data obtained in this work for the three sources using both reductions as well as the predicted photospheric fluxes are listed in Table~\ref{herscheldata}. Table~\ref{targets} lists the main stellar parameters of HIP\,103389, HIP\,107350, and HIP\,114948 (Eiroa et al., in prep.). The stellar contribution to the total SED of each star is estimated using a PHOENIX/GAIA synthetic stellar model \citep{bro05}. Specific models for $T_{\rm eff}$, \mbox{$\log g_\star$} and [Fe/H] of each star are built by interpolation from the available grid. A normalization is applied to the short wavelength range \mbox{($\lambda < 20\um$)} of the long wavelength section of the \Spitzer/IRS spectrum where available. Otherwise, the normalization is done to the flux measurements at photometric bands from $B$ to $K$. All three sources are found to have significant excess \mbox{($\ge 3\,\sigma$)} at $70\um$ and $100\um$. HIP\,114948 also has significant excess at $160\um$. HIP\,103389 and HIP\,107350 have estimated excesses at $160\um$ between $0.8\sigma$ and $2.4\sigma$, depending on the reduction. Offsets between optical positions and positions in the PACS $70\um$ images are below $1\,\sigma$ of the \Herschell pointing accuracy. In each case, the observed excess is attributed to the presence of a debris disk associated with the star. The measured FWHM of the sources (derived from a 2-D Gaussian fit) are consistent with unresolved objects. One can therefore constrain the extension of the emitting area (adopting the distance of the objects listed in Table~\ref{targets}) to a diameter of less than $7\farcs1$ ($156\,{\rm AU}$) for HIP\,103389, $7\farcs1$ ($126\,{\rm AU}$) for HIP107350, and $7\farcs2$ ($148\,{\rm AU}$) for HIP\,114948.

The spectral slopes of the three systems form Reduction~1 and Reduction~2 are listed in Table~\ref{herscheldata}. The SEDs of all three sources are found to exhibit an unusually steep decrease at the wavelength range of $70\um$ to $160\um$ (steeper than a black body radiator in the Rayleigh-Jeans regime) from the results of Reduction~1. The steepness is evaluated based on the spectral index of the SEDs between two different wavelengths (see Sect.~\ref{theory} for a detailed description). The significance of this steepness is evaluated using error propagation. However, the results from the new (but not necessarily better) version of the data reduction (Reduction~2) give a shallower decrease of the SED toward longer wavelengths. Thus, we emphasize that this is a \emph{potential} discovery of a steep decrease of the SEDs of the three sources and that a deeper understanding of the data (in particular for very faint sources) than possible until now is necessary to make a final statement. In the subsequent sections, we discuss the consequences of this steepness of the SEDs under the assumption that it is real.

\section{What is an unusually steep SED?}
\label{theory}

Infrared excesses observed from debris disk systems are the result of absorption of incident starlight by dust grains in the disk and subsequent re-emission of stellar radiation. Since the dust is colder than the star, its emission peaks at longer wavelengths (we assume that there is only one peak). In this section, a qualitative discussion is given of how steep the SED of a debris disk is expected to be. We first investigate the shape of the excess treating both the star and the disk as single temperature black body radiators. Later, we will discuss the effects of a more realistic disk model, i.e., allowing for a range of temperatures and more realistic grain properties.

\subsection{Treating the dust as a single temperature black body}

The spectral index $\Delta$ of an SED is defined as
\begin{equation}
  \Delta = \frac{\partial \log F_{\nu}}{\partial \log \nu}
\end{equation}
where $F_{\nu}$ is the total flux of the object at the frequency $\nu$. It is typically measured as the slope of the SED between two data points available, i.e.,
\begin{equation}
  \Delta_{\nu_1,\nu_2} = \frac{\log F_{\nu_2} - \log F_{\nu_1}}{\log \nu_2 - \log \nu_1}.
\end{equation}
We further denote the two frequencies $\nu_1$ and $\nu_2$ with their corresponding PACS wave bands (e.g., $\Delta_{100,70}$). The spectral index of a black body radiator amounts to \mbox{$\Delta = 2$} in the Rayleigh-Jeans regime. It is decreasing toward shorter wavelengths where the Rayleigh-Jeans approximation is not valid. A spectral index of \mbox{$\Delta > 2$} means that the SED falls off toward longer wavelengths steeper than a single temperature black body in its Rayleigh-Jeans regime. In the wavelength range where dust reemission is observed, the emission of the star can be very well approximated by a black body radiator in its Rayleigh-Jeans regime. It will only contribute as an additional component with a spectral index of \mbox{$\Delta = 2$}. Thus, the contribution of the star will -- if significant -- move the spectral slope of the whole system closer to 2.

\subsection{More exact treatment of the disk}
\label{sec_qabs}

Our first step to expand the above discussion is to allow for different dust temperatures, while the dust is still treated as a number of single temperature black bodies (i.e., a radially extended disk). A two component black body has been used successfully to model SEDs of debris disks \citep[e.g.,][]{hil08}. This results in a flatter SED \mbox{($\Delta_{\nu_1,\nu_2} < 2$)} than in the above case, since the colder dust component will contribute additional flux in the long wavelength regime of the warmer dust component. Thus, using several components with different temperatures one can only reach \mbox{$\Delta_{\nu_1,\nu_2} = 2$} in the long wavelength regime of the coldest dust component (i.e., at even longer wavelengths compared to the single black body case above). At shorter wavelengths $\Delta_{\nu_1,\nu_2}$ will always be $< 2$.

\begin{figure}
\centering
\includegraphics[width=1\linewidth]{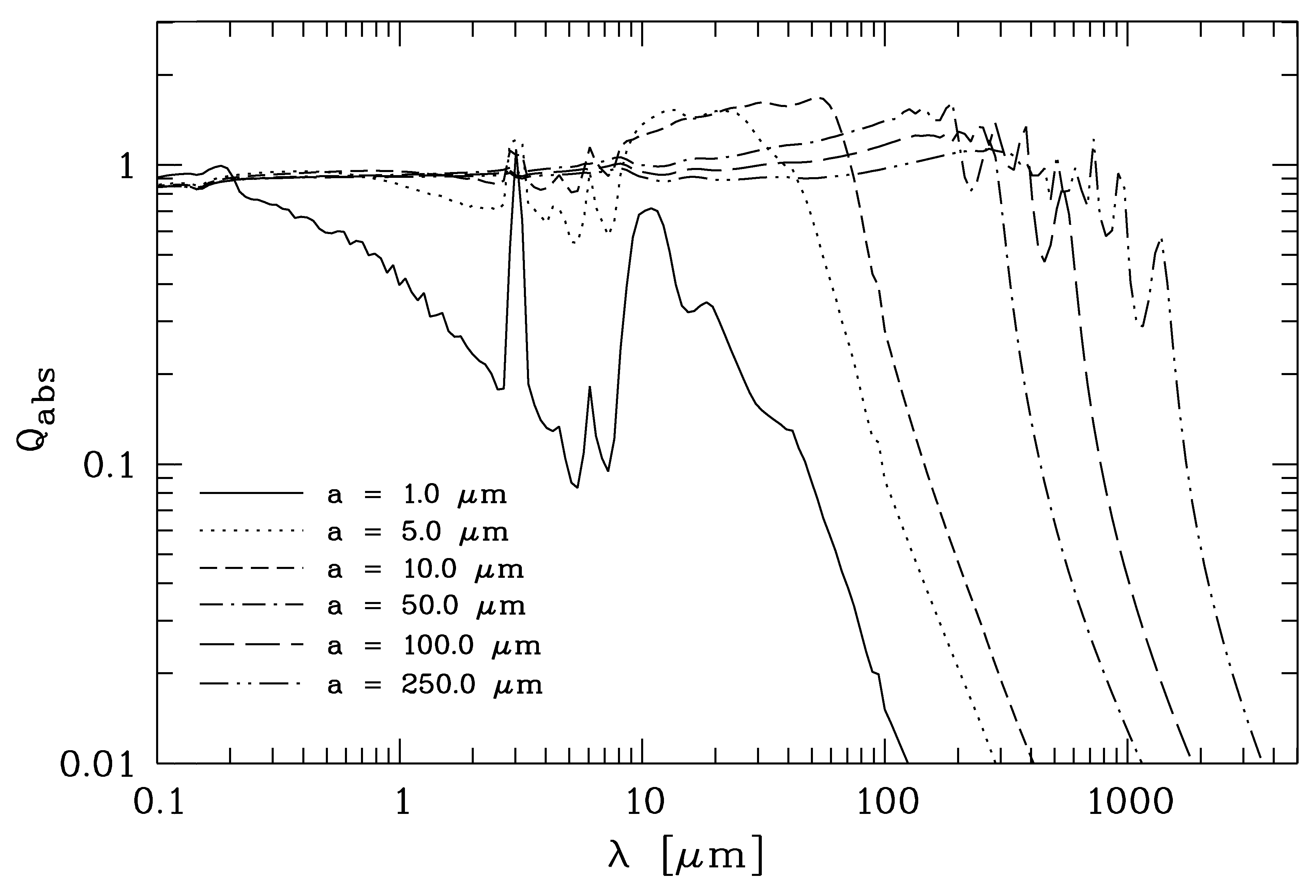}
\caption{Absorption efficiency $Q_{\rm abs}\left(\lambda\right)$ for astronomical silicate grains of different grain radii $a$ (see Sect.~\ref{sec_qabs} for details).}
\label{Qabs}%
\end{figure}

\begin{table*}
\caption{Physical properties of the three stars considered in this work (for details and references see Sect.~\ref{stelprop})}
\label{targets}
\begin{center}
\begin{tabular*}{1.0\linewidth}{l@{\extracolsep{\fill}} lll}
\hline\hline
Object                                          &  HIP\,103389            &  HIP\,107350        &  HIP\,114948         \\
\hline
Name                                            & HR\,8013, HD\,199260    & HN\,Peg, HD\,206860 & HR\,8843, HD\,219482 \\
                                                &                         &                     &                      \\
Distance (pc)                                   &  21.97                  &  17.88              &  20.54               \\
Spectral type and luminosity class              &  F7 V                   &  G0 V               &  F7 V                \\
Range of spectral types                         &  F6V$-$F8V              &  G0IV$-$V           &  F6V$-$F8V           \\
$V$, $B-V$                                      &  5.70, 0.51             &  5.96, 0.59         &  5.64, 0.52          \\ 
Absolute magnitude $M_V$, bolometric correction &  3.99, $-0.02$          &  4.70, $-0.05$      &  4.08, $-0.02$       \\
Bolometric luminosity, $L_\star$ [$L_\odot$]    &  2.025                  &  1.090              &  1.867               \\ 
Effective temperature (K)                       &  6257                   &  5952               &  6240                \\
Surface gravity, $\log g_\star$                 &  4.36                   &  4.44               &  4.31                \\
Radius, $R_\star$ [$R_\odot$]                   &  1.2                    &  0.99               &  1.17                \\
Metallicity [Fe/H]                              &  $-0.14$                &  $-0.07$            &  $-0.21$             \\
Radial velocity  (km/s)                         &  $-16.1$                &  18.1               &  0.47                \\
Rotational velocity, $v \sin i$ (km/s)          &  13.7                   &  12.8               &  9.0                 \\
Rotation period, $P$ (days)                     &  4.4 ($P/\sin i$)       &  4.7                &  --                  \\
Li{\sc ~i} equivalent width (m\AA)              &  80.0                   & 91.9                & 73.7                 \\
Space velocities $U, V, W$ (km/s)               & $-17.2$, $-10.2$, $1.1$ & $-13.9$, $-20.1$, $-11.1$ & 14.9, $-8.0$, $-5.7$     \\ 
Activity, $\log R'_{\rm HK}$                    & $-4.402$                &  $-4.48$            &  $-4.434$            \\
X-Ray luminosity, $\log L_{\rm X}/L_\star$      & $-4.71$                 &  $-4.37$            &  $-4.43$             \\
Mass, $M_\star$ [$M_\odot$]                     & 1.28,                   & 0.98                &  1.02                \\
Age (Gyr)                                       & 0.304, 0.184, 0.748($\sin i$) & 0.530, 0.162, 0.291 & 0.385, 0.115, -- \\
\hline
\end{tabular*}
\end{center}
\tablefoot{The ages given are derived using the following methods (left to right): {$\log R'_{\rm HK}$} activity index levels, ROSAT X-Ray luminosities $L_{\rm X}/L_{\rm bol}$, Rotation period. For more details and references, see Sect.~\ref {stelprop}.}
\end{table*}
Using Mie theory, one can derive absorption efficiencies $Q_{\rm abs}\left(\lambda\right)$ for more realistic dust grains. The emission of a dust grain is then described by the following equation which is a modification of Planck's law:
\begin{equation}
{B'}_{\lambda} \left(\lambda,T_{\rm dust}\right) = \frac{2hc^2}{\lambda^5} \cdot \frac{1}{\exp \left[hc/\lambda kT_{\rm dust}\right] - 1} \cdot Q_{\rm abs}\left(\lambda\right)
\end{equation}
The shape of the simple black body SED is modified by the function $Q_{\rm abs}\left(\lambda\right)$ which will result in a steeper decrease with wavelength in the Rayleigh-Jeans regime of the black body if $Q_{\rm abs}\left(\lambda\right)$ is decreasing with wavelength. Fig.~\ref{Qabs} shows $Q_{\rm abs}\left(\lambda\right)$ for astronomical silicate \citep{dra03} and different grain radii~$a$. $Q_{\rm abs}\left(\lambda\right)$ is close to 1 at short wavelengths, while it indeed exhibits a break at \mbox{$\lambda \approx 2\pi a$} and then decreases toward longer wavelengths with $\lambda^{-2}$ (or $\nu^2$). This is analog to the discussion of the opacity index, e.g., by \citet{dra06}. An SED of a dust disk that is steeper in the long wavelength regime than a black body is obviously possible. However, debris disks are usually expected to be radially extended \citep[e.g.,][]{bac09,ert11,loe11} and to contain particles of different size. This suggests a broad range of dust temperatures to be present resulting in a flattening of the SED. In addition, the grain size dependence of the break in $Q_{\rm abs}\left(\lambda\right)$ means that the grains have to be smaller than \mbox{$\approx \lambda/2\pi$} to result in an SED that falls off steeper than a black body \citep{dra06}. This suggests grains smaller than \mbox{$\sim 15\um$} for the three disks presented in this paper (Table~\ref{herscheldata}). Thus, one qualitatively expects a narrow dust belt composed of small grains to be present. While a spectral slope steeper than 2 is common for debris disks at \mbox{(sub-)mm} wavelengths, such a behavior is very unusual in the range of \mbox{$\approx 100\um$} among the few disks with characterized slope in this wavelength range \citep[e.g.,][]{hil08,roc09}. For the debris disks considered there, typical values of $\Delta$ at wavelengths around $100\um$ are in the range of 0 to 2. Note that our \Herschell observations are the first to characterize the spectral slope of a large number of debris disks at wavelengths around $100\,\um$. Earlier observations in this wavelength range suffer from low sensitivity. Thus they were limited to few bright disks or disks with very flat spectral slope in the far-ir. The completion of our survey and an analysis of the whole data set obtained (Eiroa et al., in prep.) will allow to place the three disks presented in a broader context.

We define an unusually steep SED to be an SED that exhibits a \mbox{$\Delta_{\nu_1,\nu_2} > 2$} in at least one combination of the PACS bands of $70\um$, $100\um$, and $160\um$, and for which a \mbox{$\Delta_{\nu_1,\nu_2} = 2$} would not be possible within the $1\sigma$ uncertainties. These criteria are met by the three debris disks presented in this work (Table~\ref{herscheldata}).

\section{SED modeling}
\label{modeling}

In this section we perform detailed analytical model fitting to the observed SED data to explore quantitatively the conclusions on the disk properties derived in the above qualitative discussion of the steep slope of the observed SEDs.

\subsection{Stellar properties}
\label{stelprop}
Table~\ref{targets} lists some of the main stellar parameters and other relevant observational properties of HIP\,103389, HIP\,107350, and HIP\,114948 (Eiroa et al., in prep.). These parameters were obtained using the DUNES Virtual Observatory tool\footnote{http://sdc.cab.inta-csic.es/dunes/searchform.jsp}. The spectral types are taken from the Hipparcos catalog \citep{per97, vle07}, and the range of spectral types from \citet{ski10}. Bolometric luminosities and stellar radii have been estimated from the absolute magnitudes and the bolometric corrections using the measurements by \citet{flo96}; similar values are obtained  using the bolometric correction procedure by \citet{mas06}. Effective temperatures, gravities and metallicities are mean values of spectroscopic  and photometric estimates from different works in the literature ({\it HIP\,103389} and {\it HIP\,114948}: \citealt{all99, gra06, hol09}; {\it HIP\,107350}: \citealt{val05, fuh08}). Radial velocities are taken from \citet{mal10} and \citet{kha07}. Rotational velocities of the stars are taken from \citet{hol07} and \citet{mar10}. Periods are taken from \citet{rei06} and \citet{mes03}. The Li{\sc ~i} 6707.8\,\AA~line equivalent widths are taken from \citet{mal10} in the case of HIP\,107350 and our own estimates using reduced HARPS{\footnote{Based on observations made with the European Southern Observatory telescopes obtained from the ESO/ST-ECF Science Archive Facility.}} data for the  other two stars; for these latest objects the EW(Li{\sc ~i}) have been corrected from the FeI 6707.4\,\AA~line contamination as in \citet{mal10}. All three stars show \mbox{$\log R'_{\rm HK}$} activity index levels \citep{gra06, mar10} indicating that they are active stars \citep[e.g.,][]{mal10}. The observed ROSAT X-Ray luminosities, \mbox{$L_{\rm X}/L_{\rm bol}$}, derived in the context of DUNES (Eiroa et al., in prep.) are also given.

The masses of the stars in Table~\ref{targets} are estimated from their radii and gravities. We estimate the age of the stars (Table~\ref{targets}) from the \mbox{$\log R'_{\rm HK}$} index, and periods using the relationships as given by \citet{mam08} and from the X-Ray luminosity following \citet{gar11}. For all three stars these age tracers yield consistent values of less than $\sim$~500\,Myr. We note that EW(Li{\sc ~i}) values are close to those of the Hyades envelope with an age of 600\,Myr. \citet{roc04} give an age of 370\,Myr for HIP\,107350 based on the \mbox{$\log R'_{\rm HK}$} activity index, while isochrone-based ages are 2.3\,Gyr for HIP\,103389, with a range of 0.6\,Gyr to 3.8\,Gyr \citep{hol09}, 3.1\,Gyr for HIP\,107350, with a range of 1.1\,Gyr to 4.7\,Gyr \citep{val05}, and 3.7\,Gyr for HIP\,114948, with a range of 2.1\,Gyr to 5.2\,Gyr \citep{hol09}. Given that the stars are located right on the main-sequence, stellar ages are difficult to estimate on the basis of isochrones, and that they are very sensitive to $T_{\rm eff}$ and metallicity \citep{hol09} we would favor in these cases the young values estimated from the activity tracers and rotational periods. This choice is also compatible with the kinematics of the stars and their ascription to the young moving groups, less than 1\,Gyr, of HIP\,107350 to the local association \mbox{($\sim$ 150\,Myr)} or the Hercules-Lyra \mbox{($\sim$~200\,Myr)} association \citep{mal10, lop06}, and HIP\,114948 and HIP\,103389 to the Castor Moving Group ($\sim$~200\,Myr) and the young disk population ($<$~1\,Gyr), respectively.

HIP\,107350 hosts a T dwarf companion (HN\,Peg~B, \mbox{$M = 22\,{\rm M_J}$}) with a projected physical separation of 794\,AU \mbox{($\approx 43''$)} from the host star \citep{luh07}. This is sufficiently far away, so that the presence of this companion does not affect our present analysis. A very faint source \mbox{($\approx 2\,{\rm mJy}$)} is detected in the PACS $100\,\um$ image close to the predicted position of HN\,Peg~B. However, we do not expect HN\,Peg~B to be detectable by our observations \mbox{($\approx 0.6\,{\rm mJy}$} at $8\um$). Thus, we attribute this detection to a coincidental alignment with a background source.

\subsection{Modeling the excess}
\label{fitting}

In this section, detailed analytical model fitting to the observed SED data is performed to explore quantitatively the conclusions on the disk properties derived in the above qualitative discussion. This is done using the results from Reduction~1, since these results give the steep SEDs. The fitting results are expected to differ less significantly from standard solutions using the photometric data from Reduction~2. For modeling of the three sources, we use {\tt SAnD} which is part of the DUNES modeling toolbox (\citealt{loe11}; Augereau et al., in prep.). A description of this tool can be found in the appendix. For the model fitting of the infrared excess SED, the above derived stellar properties are used. It is important to note that uncertainties of the stellar parameters as well as on the distance of the objects are not considered in our modeling of the excess. Thus, the derived uncertainties of the disk parameters are only the formal uncertainties from the fitting procedure. The uncertainties of the stellar parameters and distance would mostly affect the effective temperature and luminosity of the star and, thus, the heating of the dust -- slightly increasing or decreasing its temperature. In particular, they will not affect significantly the slope of the derived grain size distribution.

For the comparison between modeled and observed measurements we use a reduced chi-squared
\begin{equation}
\chi^{2}_{\rm red} = \frac{1}{N_{\rm dof}} \sum^{N}_{i = 1} \left(\frac{F_{i} - \tilde{F}_{i}}{\sigma_{i}}\right)^{2}~~;~~~~~N_{\rm dof} = N - N_{\rm free},
\end{equation}
where $N$ is the number of SED data points used, $F_{i}$, $\tilde{F}_{i}$, and $\sigma_{i}$ are the measured flux, the modeled flux, and the uncertainty in a data point, $N_{\rm free}$ is the number of free parameters, and $N_{\rm dof}$ is the number of degrees of freedom in our fitting. For this reduced $\chi^2$ a value of 1.0 is generally desirable in the fitting if $N_{\rm dof}$ is large, while in our case $N_{\rm dof}$ ranges from 1 to 3 where the reduced $\chi^2$ is desirable to be as small as possible.

\begin{figure*}
\centering
\includegraphics[width=1\linewidth]{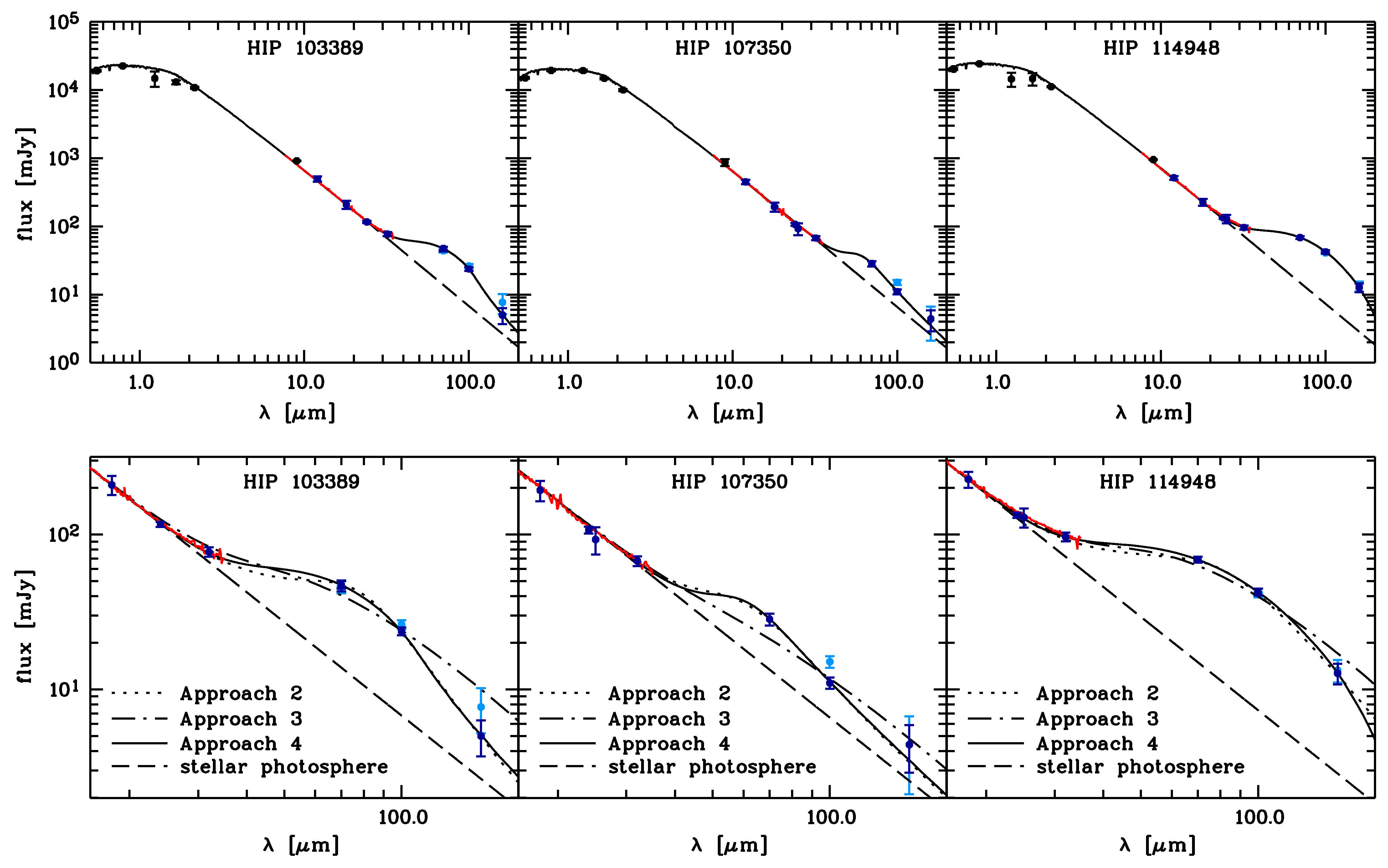} 
\caption{Observed and modeled SEDs of the three sources. The \Spitzer/IRS spectrum is plotted in red. The data points considered for the fitting are shown in dark blue. The PACS data considered are from Reduction~1, while the results from Reduction~2 are plotted in light blue for comparison. \emph{Top}: Final models from the {\tt SAnD} approaches. The modeled SEDs (solid line) for the three disks are computed from the results of Approach~4 using pure astronomical silicate (Sect.~\ref{results}). The dashed line in each panel represents the model stellar photosphere. \emph{Bottom}: Comparison between the best-fit results from the different approaches. The modeled SEDs are computed from the best-fit result of each approach using pure astronomical silicate. The results from Approach~1 are not included, since the best-fit results are found not to be in the explored range of parameters in this approach.}
\label{best-fit_sand}
\end{figure*}

Only selected data points are used. This results in a total of 8 data points included in the fitting for each source. The SEDs of the sources can be seen in Fig.~\ref{best-fit_sand}. The references for all photometric data points available for the three sources can be found in Eiroa et al. (in prep.). The selection is done by the following criteria:
\begin{itemize}
 \item All flux measurements at wavelengths \mbox{$> 10\um$} are considered, but no upper limits, since they do not give significant additional information in the present case. Measurements at these wavelengths are also included, if they are photospheric, because they exclude significant emission from the disk at these wavelengths.
 \item Measurements at wavelengths \mbox{$< 10\um$} are not included, since they are at much shorter wavelengths than the shortest wavelength at which excess is detected ($24\um \dots 32\um$).
 \item To account for the \Spitzer/IRS spectrum in a consistent way, an additional photometric point at \mbox{$\lambda = 32\um$} is extracted from these data following \citet{hil08} and is fitted along with the other photometric values. Therefore, 10 consecutive data points centered at \mbox{$\lambda = 32\um$} are averaged. The uncertainty is computed by adding in quadrature the standard deviation of these data points and a 5\% calibration uncertainty \citep[{\it Spitzer}/IRS instrument handbook v4.0;][]{tep11}. This will prove to be sufficient to get the whole \Spitzer/IRS spectrum properly reproduced by all the best-fit models.
\end{itemize}

\begin{table*}
\caption{Explored parameter space for the fitting based on simulated thermal annealing (for details see Sect.~\ref{fitting})}
\label{paratab_sand}
\begin{center}
\begin{tabular*}{1.0\linewidth}{c@{\extracolsep{\fill}} ccccccccc}
\hline\hline                 
   & \multicolumn{2}{c}{Approach~1} & \multicolumn{2}{c}{Approach~2} & \multicolumn{2}{c}{Approach~3} & \multicolumn{2}{c}{Approach~4} & \\
  \cline{2-3} \cline{4-5} \cline{6-7}  \cline{8-9}
  Parameter &  Range & \# values & Range & \# values & Range & \# values & Range & \# values & distribution \\ 
\hline
  $R_{\rm{in}}$ [AU]            & 3 -- 100      & 817   & 3 -- 100      & 817   & 3 -- 100      & 817   & 3 -- 100      & 817   & temp        \\
  $R_{\rm{out}}$ [AU]           & 5 -- 300      & 671   & 5 -- 300      & 671   & 5 -- 300      & 671   & 5 -- 300      & 671   & temp        \\
  $\alpha$                      &$-2.0$ -- $3.5$& 56    & fixed         & 1     & fixed         & 1     & fixed         & 1     & linear      \\
  $a_{\rm{min}}$ [$\um$]        & 0.2 -- 20.0   & 448   & 0.2 -- 20.0   & 448   & 0.2 -- 20.0   & 448   & 0.2 -- 20.0   & 448   & logarithmic \\
  $a_{\rm{max}}$ [$\um$]        & fixed         & 1     & fixed         & 1     & fixed         & 1     & 1.0 -- 1000.0 & 674   & logarithmic \\
  $\gamma$                      & 2.0 -- 5.0    & 31    & 2.0 -- 10.0   & 81    & fixed         & 1     & fixed         & 1     & linear      \\
  $M_{\rm{dust}}$ [$M_{\oplus}$] & free          & \dots & free          & \dots & free          & \dots & free          & \dots & continuous  \\
  Composition                   & 0\%, 50\% ice & 2     & 0\%, 50\% ice & 2     & 0\%, 50\% ice & 2     & 0\%, 50\% ice & 2     & \dots       \\
\hline
\end{tabular*}
\end{center}
\end{table*}

The capabilities of {\tt SAnD} are used to explore a very broad range of parameters. The following model is employed:
\begin{itemize}
 \item Radial surface density distribution \mbox{$\Sigma\left(r\right) \propto r^{-\alpha}$} with inner and outer cut-off radii $r_{\rm in}$ and $r_{\rm out}$,
 \item Differential grain size distribution $dn(a) \propto a^{-\gamma} da$ with lower and upper cut-off size $a_{\rm min}$ and $a_{\rm max}$,
 \item Two possible grain compositions -- pure astronomical silicate \citep{dra03} and a 1:1 mixture of astronomical silicate and ice (\citealt{loe11}; Augereau et al., in prep.) -- to explore the possibility that water ice might be a significant constituent of debris disk dust and might be responsible for the peculiar shape of the SEDs.
\end{itemize}

Since the number of free parameters in the fitting is close to the number of data points used, the fitting is expected to be very degenerate. To explore the parameter space in an efficient way and to find parameters that can be fixed, because they have a unique best-fit result or they have no significant effect on the fitting at all, we use different approaches. These approaches are described below. The parameter space explored for each approach is listed in Table~\ref{paratab_sand}. The fitting results are compiled in Table~\ref{results_sand}\addtocounter{table}{1}. Simulated SEDs from our best-fit models are shown in Fig.~\ref{best-fit_sand}. The parameter space explored and the parameters fixed in each approach are motivated by the results of the previous approaches as described in the following.

\subsubsection{Approach~1: Exploring the parameter space}

In a first approach, a range of free parameters is considered that includes those used to explain most of the known debris disks. An upper grain size of 1\,mm is chosen, large enough to consider any effect of large grains on the fluxes at all wavelengths observed. Seven free parameters are used ($r_{\rm in}$, $r_{\rm out}$, $\alpha$, $a_{\rm min}$, $\gamma$, $M_{\rm dust}$, and dust chemical composition). This is the largest number possible considering a total of 8 SED measurements. Although strong degeneracies are expected, this approach is used to explore the parameter space without strong initial constraints that might bias the results. Parameters that have a unique solution or that have no significant effect on the fit can be fixed in subsequent approaches.

A very narrow ring structure is found to be the best-fit for all three disks, although the parameters describing the spatial dust distribution are not constrained very well. Furthermore, a very steep grain size distribution with a value for the exponent $\gamma$ very close to the edge of the explored parameter space (5.0) is found as best-fit for all three SEDs (Table~\ref{results_sand}). In the following, it is referred to this as an underabundance of large grains. Since this is an atypical, but not unexpected result (Sect.~\ref{theory}), more effort is put on the evaluation of the significance of this result in the subsequent approaches. As the radial extent of the disks has been found to be narrow, the exponent of the radial surface density distribution is fixed in the further fits to \mbox{$\alpha = 0.0$} (constant surface density), decreasing the explored parameter space by one dimension. In a narrow ring, this will not have any effect on the SED.

\subsubsection{Approach~2: A larger range of possible values for $\gamma$}

Now, the aim is to obtain a best-fit value for $\gamma$ that is included in the considered parameter space. A range of possible values for $\gamma$ of $2.0$ to $10.0$ is explored. All other parameters (beside $\alpha$ which is now fixed as described above) have the same ranges as in Approach~1. This results in a total of 6 free parameters.

For HIP\,107350, the range of explored values of $\gamma$ seems to be still too small. This is ignored, since such a very large value is not significant (confidence levels: 5.9 \dots 10.0, Table~\ref{results_sand}). For the other two objects, the value of $\gamma$ is well within the explored range. The confidence levels of the inner and outer radius suggest that the ring-like shape of the disk is not very significant, even for a constant surface density.

\subsubsection{Approach~3: Fixing $\gamma$ to 3.5}

Fixing the value of $\gamma$ to 3.5 and having a fixed upper grain size of \mbox{$a_{\rm{max}} = 1.0\,{\rm mm}$}, one can force a fit where the parameters of the grain size distribution are consistent with an equilibrium collisional cascade \citep{doh69}. The explored ranges of all other parameters are unchanged compared to Approach~2. This results in a total of 5 free parameters.

The resulting $\chi^2$ is much worse (by a factor of 2.5 to 12.6, Table~\ref{results_sand}) compared to the results from Approach~2. The changes of the radial position of the dust ring are in line with the expectations from the much lower abundance of small (warm) particles. 

\subsubsection{Approach~4: Fixed value of \mbox{$\gamma = 3.5$}, but free upper grain size}

Another possibility to produce an underabundance of large grains in the model is to let the upper grain size be a free parameter. Since a very steep size distribution and a small upper grain size have comparable results (removing large grains from the model), one can fix the value of $\gamma$ to $3.5$ to be consistent with an equilibrium collisional cascade. The resulting upper grain size is expected to be sufficiently small. This results in a total of 6 free parameters.

This approach gives in general the best $\chi^2$ (Table~\ref{results_sand}). Upper grain sizes of few tens of micron are found. The low radial extend of the debris ring is again not very significant, but still the best-fit.

\subsubsection{Grid search and inclusion of spatial information for HIP\,114948}

From our discussion in Sect.~\ref{theory} we are not able to constrain our parameter space by any reasonable physical assumptions. Since {\tt SAnD} is designed to explore a large, high-dimensional parameter space using a statistical approach, it was particularly useful to find the global best-fit to the data in the above approaches. The over-all result of our fitting (a lack of large grains) is clearly visible for HIP\,103389 and HIP\,107350. For HIP\,114948 the result is not that stringent. The value of \mbox{$\gamma = 4.7^{+0.5}_{-0.3}$} from Approach~2 of the {\tt SAnD} fitting is significantly smaller than the values derived from the other two disks \citep[although still significantly larger than the value of 3.5 expected from a standard equilibrium collisional cascade,][]{doh69}. Thus, we perform a grid search of the parameter space using GRaTer \citep{aug99,leb11} to confirm our results from the statistical approach of {\tt SAnD}. Furthermore, the constraint that the source is spatially unresolved is included in this fitting. This is not expected to significantly change the best-fit results (since these results are consistent with an unresolved disk) but to further constrain the range of values possible within the uncertainties of the derived values (since very extended configurations that would clearly be resolved are also included in the confidence levels from pure SED fitting). The results from fitting with and without the spatial constraints are compared.

Since the general result from the {\tt SAnD} fitting (steep grain size distribution, large lower grain size) is independent from the approach used, we decide to use the standard approach of GRaTer. This is a power-law approach \mbox{($d n\left(a\right) \propto a^{-\gamma} da$)} with upper and lower limit for the differential grain size distribution and a two-power-law approach for the radial surface density distribution:
\begin{equation}
 \Sigma(R) \propto \sqrt{2\left(\left(R/R_0\right)^{-2\alpha_{\rm in}} + \left(R/R_0\right)^{2\alpha}\right)^{-1}},
\end{equation}
where $\alpha$ and $\alpha_{\rm in}$ are slopes of the radial surface density distribution outside and inside a peak position. We fix \mbox{$\alpha_{\rm in} = 10$} to get a sharp inner edge close to $R_0$ comparable to the single power-law approach used for the {\tt SAnD} fitting.

\begin{figure}
\centering
\includegraphics[width=\linewidth]{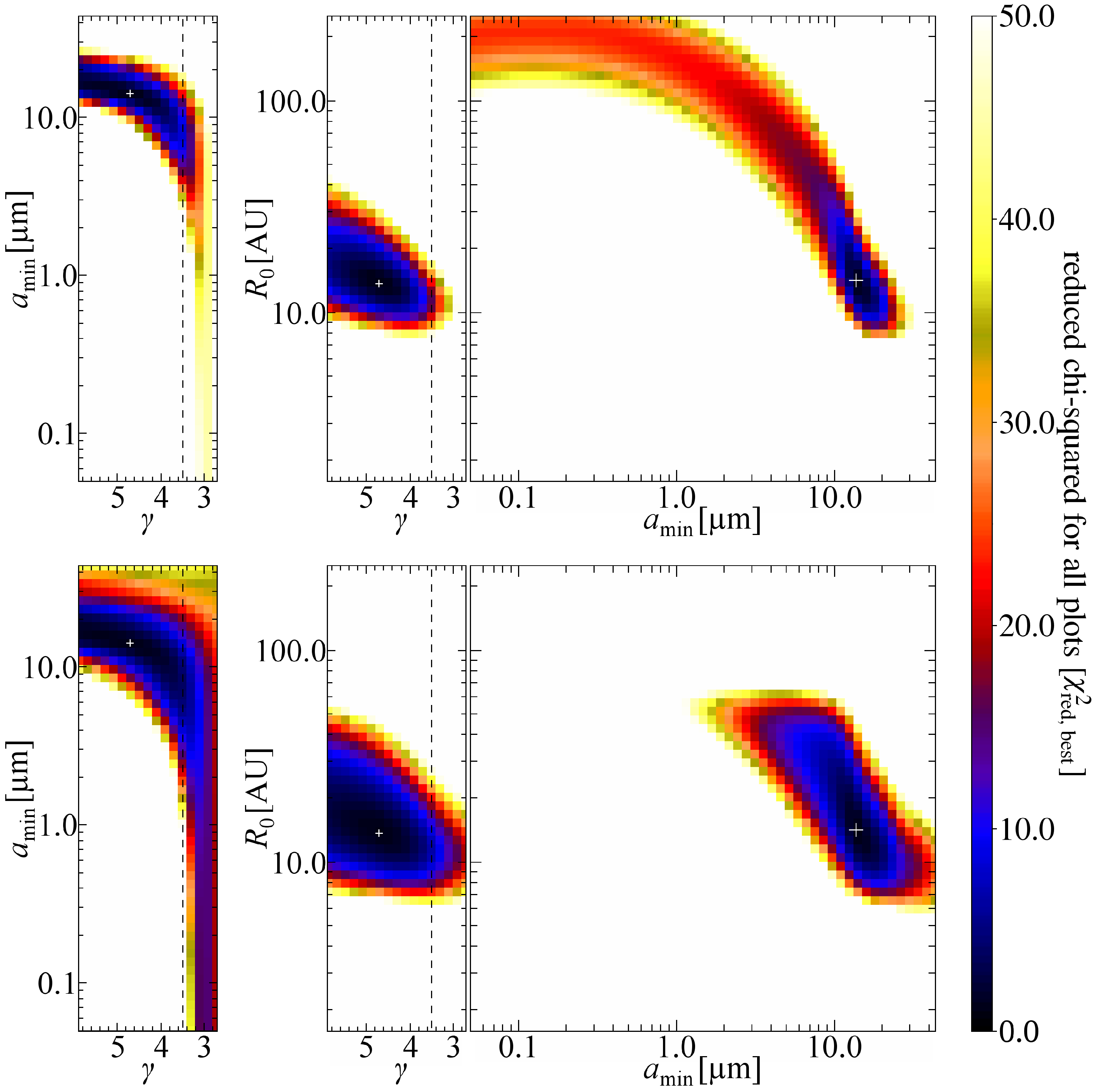}
\caption{$\chi^2$ maps for HIP\,114948. The maps are created as two dimensional cuts through the $\chi^2$ distribution in the parameter space searched with GRaTer without (\emph{top}) and with (\emph{bottom}) the spatial information included. The parameters not plotted are fixed to their best-fit value, respectively (Table~\ref{grater_fit}). The correlations between the three most relevant parameters ($\gamma$, $a_{\rm min}$, and $R_0$) are shown. The white cross denotes the position of the best-fit (lowest $\chi^2$). The dashed vertical line denotes the value of \mbox{$\gamma = 3.5$} expected from a standard equilibrium collisional cascade.}
\label{maps_chi2}%
\end{figure}

The spatial constraints are included in our fitting using a radial brightness profile derived by radial averaging of the sources image in our $100\um$ data. After a Gaussian is fitted to the image to derive the source center, the average flux and distance from the center is derived for all pixels in one pixel wide radial bins. The uncertainty on the flux in each bin is derived using the standard deviation of the pixel values in each radial bin. Images are simulated from the models, convolved with the PACS PSF measured on the bright standard star $\alpha$\,Boo and profiles are extracted. These profiles are compared to the one derived from the data. The same number of SED data points as before is considered. In addition, a total of 10 profile points are used. An over-all reduced $\chi^2$ including SED and profile data is computed following Eq.~\ref{eq:chi2} (using only one profile). The explored parameter space and best-fit values for the fit with and without the profile included are given in Table~\ref{grater_fit}. These results should be compared to the results from Approach~2 of the {\tt SAnD} fitting, since these two approaches are the most similar. They are found to be consistent. Maps of the distribution of the reduced $\chi^2$ in the parameter space for the fits including and not including the profile have been produced as cuts through the parameter space in three planes. The remaining parameters are fixed to their best-fit values, respectively. These maps are displayed in Fig.~\ref{maps_chi2}. We find that the best-fit values do not change significantly when including the spatial information. This confirms that the results from the pure SED fitting are fully consistent with a spatially unresolved disk. Fig.~\ref{maps_chi2} illustrates a number of results from the mapping of the parameter space:

\begin{table*}
\caption{Fitting result for HIP\,114948 using the grid search method with GRaTer}
\label{grater_fit}
\begin{center}
\begin{tabular*}{1.0\linewidth}{c@{\extracolsep{\fill}} ccccccc}
\hline\hline                 
 Parameter                                & Range        & Spacing     & \# values & Fit w/o spatial constraints & Fit w/ spatial constraints \\
\hline
 $R_0$ [AU]                               & 1.0 -- 250.0 & logarithmic & 60        & 13.7 & 13.7 \\
 $\alpha$                                 & 0.0 -- 10.0  & linear      & 11        & 8.0  & 9.0  \\
 $a_{\rm{min}}$ [$\um$]                   & 0.05 -- 43.5 & logarithmic & 55        & 14.1 & 14.1 \\
 $\gamma$                                 & 2.7 -- 5.9   & linear      & 17        & 4.7  & 4.7  \\
 $V_{\rm ice}/(V_{\rm si} + V_{\rm ice})$ & 0.0 -- 0.9   & linear      & 10        & 0.6  & 0.6  \\
 $M_{\rm dust}$ [$M_\oplus$]              & $1.0\times10^{-6}$ -- $4.0\times10^{-3}$ & logarithmic & 100 & $2.36\times10^{-5}$ & $2.20\times10^{-5}$ \\
 reduced $\chi^2$                         & --           & --          & --        & 0.19 & 0.12  \\
\hline
\end{tabular*}
\end{center}
\end{table*}

\begin{itemize}
 \item There is a correlation between the lower dust grain size $a_{\rm min}$ and the exponent $\gamma$ of the size distribution. However, from our SED fitting one can rule out that any combination of significantly smaller grains and smaller exponent of the size distribution results in reasonable fits to the data (which would be closer to the expectations from a collisional cascade and grains close to the blow-out size of the system of \mbox{$a < 1.0\um$}). 
 \item There is nearly no correlation between the radial position of the dust belt and the exponent of the grain size distribution in both the case without and with the spatial information included.
 \item The degeneracy between the lower dust grain size and the radial position of the dust belt known from SED fitting is clearly visible. It is also obvious that the inclusion of the spatial information in part breaks this degeneracy since it excludes a large \mbox{($\geqslant 70\,{\rm AU}$)} radial distance of the belt from the central star.
 \item In particular, the results of $a_{\rm min}$ and $\gamma$ seem to be less constrained when including the spatial information. This is in part because the spatial information gives no constraints on the parameters of the grain size distribution (beside the partly broken degeneracy of the lower dust grain size and the radial position of the dust which results in an indirect constraint on $a_{\rm min}$). Thus, including the radial profile we include a number of measurements that are not sensitive to changes in $\gamma$, which results in a less significant increase of the reduced $\chi^2$ with changes of $\gamma$ in this case. In addition, the data points in the profile can not be considered to be independent from each other. Although this is mitigated in part by the weighting in Eqs.~\ref{eq:chi2} to~\ref{eq:chi2prof}, the combination of $\chi^2$ derived from SED and profile is not fully consistent, in particular in the case of unresolved sources. To measure the quantity and uncertainty of a model parameter through model fitting using reduced $\chi^2$, one should in general only include measurements in the fitting process that are sensitive to the parameter (as far as one can tell in advance) and independent from each other. Thus, the uncertainties on $\gamma$ derived from pure SED fitting should be taken more reliable here.
\end{itemize}

\section{Results}
\label{results}

The general results and conclusions that can be drawn from the fitting approaches are summarized in the following.
\begin{itemize}
 \item Values of $\gamma = 4.7$ to $10.0$ or a small upper grain size are found rather than the expected value of $\gamma \approx 3.5$ and a large $a_{\rm max}$. Models with \mbox{$\gamma = 3.5$} do not reproduce the observed SEDs in a reasonable way. This result is not strongly affected by any modeling degeneracies found (that are included in the estimate of the uncertainties).
 \item Evidence is found for a large lower grain size of the dust compared to the expected blow-out size of the systems ($0.5\um$ to $0.9\um$). However, the lower grain size depends very much on the approach used. Earlier studies of debris disk SEDs had similar results \citep[e.g., HD\,107146;][]{roc09} that have been at least mitigated and partially attributed to modeling degeneracies in pure SED fitting after including resolved data in the fitting process \citep{ert11}. Thus, this result has to be treated with caution.
 \item The dust in all three disks must be located at a distance from the star of no more than a few tens of AU as can be concluded from the modeling results and the fact that the disks are spatially unresolved in the PACS images. Stronger constraints on this value are not possible due to degeneracies in the modeling.
 \item The disks appear to be narrow rings. However, the uncertainties -- in particular of the outer disk radius -- are very large and also very broad disks can result in fits on the SED that fall within the derived $3\,\sigma$ confidence levels. This can only partly be excluded by the additional constraint that the disks are spatially unresolved.
\end{itemize}
It is important to note that these results depend in part on the model used and on the parameter space explored. For example, the composition and shape of the dust grains have not been explored in detail in our modeling, but might allow one to reproduce the SEDs with models that are more in line with a standard equilibrium collisional cascade.

Furthermore, the results from Reduction~2 have not been considered in the fitting. Considering these photometric measurements and uncertainties instead of those from Reduction~1 would lower the significance of the underabundance of large grains. As a result, values are expected to be found for the model parameters that are more in line with those expected from an equilibrium collisional cascade. Without a deeper understanding of the data obtained with \Herschel, it is not possible to make a final statement about the significance of our modeling results.

Each approach presented results in a different set of best-fit parameters and a different $\chi^2$, while the results can be interpreted in a consistent way. To find from the fitting results a final model of each disk, the following selection is applied:
\begin{itemize}
 \item The results from Approach~1 have to be ruled out, because the best-fit parameters from this approach are close to the edge of the explored parameter space.
 \item In general, the SEDs cannot be fitted by a grain size distribution that includes a significant amount of large grains (Approach~3).
 \item The results from Approach~2 (free $\gamma$, \mbox{$a_{\rm max} = 1.0\,{\rm mm}$}) and Approach~4 (free $a_{\rm max}$, \mbox{$\gamma = 3.5$}) are consistent, while the results from Approach~4 and pure silicate represent in general the best fit to the data. Thus, these results are considered as the best-fit models from the fitting.
\end{itemize}
From these final models, one finds dominating dust temperatures (the temperature of the smallest grains that dominate the emission) of 63\,K, 50\,K, and 73\,K for HIP\,103389, HIP\,107350, and HIP\,114948.

\subsection{The origin of the dust}

The existence of micron-sized dust grains around main sequence stars is usually explained by collisions of larger bodies such as planetesimals \citep[e.g.,][]{kri10}. In such a scenario one would expect a significant amount of larger grains (several tens of micron to millimeter sized particles) to be present that are produced by the collisional cascade also producing the smallest grains \citep{doh69}. These grains significantly contribute to the SED in the observed wavelength regime. Assuming that the steep slopes of the SEDs are real, we find from our modeling strong evidence for an underabundance of these larger grains. To our knowledge, these are the first debris disks discovered, that potentially exhibit such a peculiar shape of the SED in the wavelength regime of $70\um$ to $160\um$. All this would suggest that these objects are exceptions from the common understanding of dust creation in debris disks. On the other hand, six more disks with similar shape of the SED have recently been identified in our DUNES survey, which might suggest that this phenomenon is common among low-mass (i.e., low luminosity, Tables~\ref{results_sand} and~\ref{properties}) debris disks\footnote{For an overview of the DUNES results we refer to Eiroa et al., in prep.}. The fact that they are very faint (Table~\ref{properties}) might imply that we are faced with a new class of debris disks that were not revealed earlier due to limited sensitivity in the relevant wavelength range. In the following, we discuss scenarios that might be capable to explain the presence of a disk as it has been modeled for the three systems.

\smallskip
\noindent{\bf Scenario~1: Significant deviation from the conditions required for a standard equilibrium collisional cascade}\\
For the standard equilibrium collisional cascade, a number of assumptions are made that are not necessarily valid in debris disks due to the effects of radiation pressure and Poynting-Robertson (PR) drag (Sect.~\ref{intro}). For the massive debris disks known so far, PR drag is expected to be negligible \citep{wya05}. Following \citet{bac93}, one can estimate the ratio between collisional time scale and PR time scale for the three disks to $10^{-2} \dots 10^{-3}$. However, the disk models used (best-fit from Approach~4, pure silicate) represent very narrow rings, which is not a significant fitting result. Assuming a ring with a width of 10\,AU starting at the inner radius found from the fitting, one finds $t_{\rm coll}/t_{\rm PR} = 3\times10^{-2} \dots 5\times10^{-2}$. It is not possible to put strong constraints on the dust dynamics due to the crude estimate of the time scales. Thus, transport mechanisms may play a role in the dust dynamics of the small grains in these disks. The effects of these processes have been modeled, e.g., by \citet{kri06,the07,wya11}. In particular, radiation pressure leads to a wavy structure of the grain size distribution increasing the slope of the grain size distribution in certain ranges. PR drag results in a depletion of small grains, increasing the dominating grain size, but mitigating the wavy structure of the size distribution.

\begin{table*}
\caption{Simulated observational properties of the disks derived from our final models}
\label{properties}
\begin{center}
\begin{tabular*}{1.0\linewidth}{c@{\extracolsep{\fill}} ccccccccc}
\hline\hline                 
   Source & $L_{\rm dust}/L_{\star}$ & \multicolumn{2}{c}{$\left(F_{\rm dust}/F_{\star}\right)_{\rm 0.6\um}$} & \multicolumn{2}{c}{$\left(F_{\rm dust}/F_{\star}\right)_{\rm 0.8\um}$} & \multicolumn{2}{c}{$\left(F_{\rm dust}/F_{\star}\right)_{\rm 1.1\um}$} & \multicolumn{2}{c}{$\left(F_{\rm dust}/F_{\star}\right)_{\rm 2.2\um}$} \\
  \cline{3-4} \cline{5-6} \cline{7-8} \cline{9-10}
   & & face-on & edge-on & face-on & edge-on & face-on & edge-on & face-on & edge-on \\
\hline
 HIP\,103389 & $1.5 \times 10^{-5}$ & $1.1 \times 10^{-6}$ & $1.9 \times 10^{-4}$ & $1.1 \times 10^{-6}$ & $1.7 \times 10^{-4}$ & $1.3 \times 10^{-6}$ & $1.8 \times 10^{-4}$ & $1.8 \times 10^{-6}$ & $1.5 \times 10^{-4}$ \\
 HIP\,107350 & $0.6 \times 10^{-5}$ & $4.9 \times 10^{-7}$ & $7.4 \times 10^{-6}$ & $4.6 \times 10^{-7}$ & $8.5 \times 10^{-6}$ & $5.5 \times 10^{-7}$ & $1.3 \times 10^{-5}$ & $6.9 \times 10^{-7}$ & $3.9 \times 10^{-5}$ \\
 HIP\,114948 & $2.5 \times 10^{-5}$ & $2.1 \times 10^{-6}$ & $3.4 \times 10^{-4}$ & $2.0 \times 10^{-6}$ & $3.0 \times 10^{-4}$ & $2.6 \times 10^{-6}$ & $3.3 \times 10^{-4}$ & $3.7 \times 10^{-6}$ & $2.6 \times 10^{-4}$ \\
\hline
\end{tabular*}
\end{center}
\end{table*}

\smallskip
\noindent{\bf Scenario~2: Different chemical composition or physical shape of the dust grains than the assumed one}\\
One might imagine a dust composition emitting significantly more efficiently than astronomical silicate or ice in the wavelength range of $70\um$ to $100\um$ compared to longer wavelengths. Therefore, one would need grains with a break in $Q_{\rm abs}\left(\lambda\right)$ (Sect.~\ref{theory}) at shorter wavelengths (in particular for large grains). Also, if the large grains were significantly colder than expected, their emission would be reduced. This would both allow a significantly larger amount of large grains to be present in the systems than modeled. Furthermore, Mie calculations might result in an incorrect approximation of $Q_{\rm abs}$ particularly for large grains (tens of microns or larger), e.g., due to incorrect assumptions on the physical shape of the grains.

\citet{vos06} simulated the effect of porosity on the absorption efficiency of dust grains. They found that the temperature of dust grains at a given distance from a star decreases significantly with increasing porosity. If one now has small, compact grains and larger, porous grains, the difference in temperatures of the smaller and larger grains can be increased significantly. This would be the case, if the large grains producing the smaller ones through collisions were porous, composed of smaller, compact units in the order of the lower grain size derived from the modeling of the three systems (i.e., $5\um$ to $10\um$). Such a scenario would result in an overabundance of grains of this size compared to the larger, weaker grains. Such an overabundance would shift the lower cut-off of our adopted power-law distribution toward this grain size, although smaller grains are present in the system.

\smallskip
\noindent{\bf Scenario~3: A shepherding planet}\\
Planets can trap dust particles into mean-motion resonances \citep[e.g.,][]{wya06}. This results in a barrier against particles moving inward due to PR drag. This barrier is less efficient for very small grains for which PR drag is very strong \citep{rei11}. On the other hand, very large grains and planetesimals are not significantly affected by PR drag.

The dust seen in these disks might be produced in a faint, transport dominated \citep{kri10} debris disk further away from the star, too cold and too faint to be detected in the available data. The small particles (few tens of micron and smaller) would then be dragged inward by PR drag. A possible planet present further inward in the system should trap the particles into resonance. This would result in an accumulation of particles with a very distinct range of sizes, which would explain both the lack of large grains (at least in an abundance and at a position where they would be detectable) and the lack of small grains. Such a ``dust trail'' has been observed, e.g., to be associated with our Earth and has been modeled to predominantly consist of grains of \mbox{$\approx 12\um$} and larger \citep{der94}. It will have to be evaluated, whether an outer planetesimal belt capable to produce enough dust through collisions to replenish the dust in the disk can be faint enough not to be detected by the \Herschell observations, and under which conditions the planet can trap enough dust into resonance to produce the disk observed. This has to be done through detailed dynamical and collisional modeling \citep[e.g.,][]{rei11}, which is not within the scope of the present work.

\subsection{Observational perspectives}

Further observations with present and near future instruments can help to reasonably increase our understanding of these debris disks. In the following, we discuss a number of observations that might result in very valuable data.

\begin{itemize}
 \item Since the most extended models of the three disks predict radial extents of only \mbox{$\approx 2\farcs2$}, these objects are at the edge of what is resolvable with \emph{coronagraphy}. Furthermore, the disks are very faint in scattered light (Table~\ref{properties}). It is important to note that in the case of edge-on orientation most of the flux comes from forward scattering and will then be concentrated close to the star (projected physical separation) with no contribution to the signal in coronagraphic observations. A contrast ratio of \mbox{$< 10^{-6}$} is not accessible to present instruments.
 \item Successful \emph{optical/near-infrared imaging of planetary companions} would give strong evidence that Scenario~3 (shepherding planet) is responsible for the peculiarities of the disk. Determining the position of this planet would also help to further constrain the position of the dust in this scenario. The youth of the stars means good chances to directly image giant planets at separations of \mbox{$\geqslant 1''$} from the star with present methods \citep[e.g.,][]{marois10}.
 \item The predicted extents of the disks are in a range easily resolvable with \emph{ALMA}. However, the disks are already very faint at PACS wavelengths and the expected surface brightness is decreasing particularly steep toward wavelengths accessible with ALMA. It is not clear, whether ALMA observations are sensitive enough to detect these disks. On the other hand, at \mbox{(sub-)mm} wavelengths even upper limits in the sensitivity range reachable are expected to provide useful further constraints on the shape of the SEDs. From the results of our fitting Approach~4, predicted fluxes of star and disk at $350\um$ (the shortest wave band accessible with ALMA) are $0.55\,{\rm mJy}$ and $0.11\,{\rm mJy}$ for HIP\,103389, $0.53\,{\rm mJy}$ and $0.05\,{\rm mJy}$ for HIP\,107350, and $0.60\,{\rm mJy}$ and $0.20\,{\rm mJy}$ for HIP\,114948, respectively.
 \item Further \emph{photometry and spectroscopy in thermal reemission} would help to constrain the radial distribution of the dust and the size distribution of the small grains (wavelengths between $40\um$ and $70\um$) and the steep slope and the  shape of the SEDs at longer wavelengths \mbox{($\lambda > 70\um$)}. The Stratospheric Observatory for Infrared Astronomy (SOFIA) is most promising to provide observational capabilities in the relevant wavelength regime in the near future.
\end{itemize}
\section{Conclusions}
\label{conc}

The first data at wavelengths \mbox{$> 70\um$} for the three debris disks discussed in this paper have been presented and modeled. All three sources potentially exhibit an unusually steep decrease of the SED in the wavelength range between $70\um$ and $160\um$. In a general discussion, it has been shown that this peculiar shape of the SED is an indicator for a deviation from the case of a standard equilibrium collisional cascade assuming standard grain composition (astronomical silicate or ice) and shape (compact, spherical grains). However, it has also been shown that the understanding of faint source photometry obtained with \Herschell is still incomplete and that the results presented here depend very much on the actual version of the data reduction pipeline. Provided that the steep decrease of the SEDs is real, modeling implies that the thermal emission from these disks is dominated by a very distinct grain size regime of several micron to few tens of micron. The disks have been modeled as narrow rings with a significant underabundance of large grains. A number of possible explanations for such an unusually steep shape of the SEDs have been discussed. Six more candidates for this new class of debris disks have been identified so far from the ongoing DUNES survey. This is the first published discovery of debris disks that potentially exhibit such a peculiar shape of the SED in this wavelength regime.

\begin{acknowledgements}
We thank Kate Su for the re-reduction of the Spitzer/MIPS photometry used in this work. Furthermore, we thank the whole DUNES team for valuable discussion. S. Ertel thanks for financial support from DFG under contract WO\,857/7-1 and for general support from K. Ertel. C. Eiroa, J. Maldonado, J.~P. Marshall, and B. Montesinos are partially supported by Spanish grant AYA 2008/01727. J.-C. Augereau and J. Lebreton thank financial support through PNP-CNES. A.~V. Krivov ans T. L\"ohne thank for financial support from DFG under contracts KR\,2164/9-1 and LO\,1715/1-1. O. Absil is supported by an F.R,S.-FNRS Postdoctoral Fellowship. S. Ertel, J.-C. Augereau and J. Lebreton thank the French National Research Agency (ANR) for financial support through contract ANR-2010 BLAN-0505-01 (EXOZODI). This work was partly funded by the Funda\c{c}\~ao para a Ci\^encia e a Tecnologia (FCT) through the project PEst-OE/EEI/UI0066/2011.
\end{acknowledgements}

\bibliographystyle{aa}

\bibliography{../../../bibtex}

\begin{appendix}

\section{Description of SAnD}

When performing analytical model fitting of debris disks, one is faced with a number of challenges. On the one hand, recent results \citep[e.g.,][]{bac09,ert11} reveal more and more the complexity of the known debris disks and the inability to reproduce the available data with simple models. Furthermore, standard assumptions (e.g., lower grain size consistent with the blow-out size of the system, grain size distribution exponent of $-3.5$) have been proven to be inadequate (e.g., \citealt{ert11}; this work), and unexpected results like an outward increasing surface density distribution \citep{loe11} have been found. The increasing number and quality of the data available (in particular, spatially resolved images from {\it Herschel} and ALMA) allow one to break modeling degeneracies through simultaneous multi-wavelength modeling and to increase the complexity of the models. On the other hand, finding a best-fit model in a complex parameter space results in significant challenges on the fitting method and the simulation of the observed data in order to reduce the computational effort necessary.

To rise to these challenges we, developed the tool {\tt SAnD}. It is able to fit SED data and radial profiles in thermal reemission simultaneously using a simulated annealing approach \citep{pre92} on a grid of possible values in the parameter space. It uses an analytical approach for the radial density distribution of a rotationally symmetrical disk as well as for the grain size distribution. The SED is computed in an analogous way to \citet{wol05}. Resolved images are computed with sufficient resolution and convolved with the telescope PSF. Radial profiles are then extracted along the major and/or minor axis. Each image (i.e., each pair of major and minor axis profiles or single major axis profile, if no minor axis profile is extracted, e.g., in azimuthally symmetrical disk images) is scaled by a factor $x_i$ to minimize the $\chi^2_{i,\rm prof}$ (Eq.~\ref{eq:chi2prof}). This way, only the shape of a radial profile is fitted. This is done because flux calibration uncertainties have to be considered only once per wavelength, but would be included for each data point in the radial profiles if absolute profiles were considered. Furthermore, deficits of our models to reproduce the absolute flux at one wavelength at which also radial profiles are fitted (e.g., due to uncertainties in the optical properties of the dust) would be considered in each profile data point and the SED data point which would result in an over-weighting of the data at this wavelength. Simulated and observed multi-wavelength data are then compared using the following reduced $\chi^{2}$:
\begin{eqnarray}
  \chi^2_{\mathrm{r}} & = & \frac{N\dma{tot}\left(\sum_{i=0}^{M} w_i\right)^{-1}}{N\dma{tot} -
    N\dma{free}}  \left(w_0 \frac{\chi^2\dma{SED}}{N_0} +
    \sum_{i=1}^{M}\limits w_i
    \frac{\chi^2_{i,\mathrm{prof}}}{N_{i,1}+N_{i,2}}\right) ,
  \label{eq:chi2} \\ \smallskip
   \textrm{with} & & \chi^2\dma{SED} = \sum_{k=1}^{N_0}\limits \left(\frac{F_k - 
      \tilde{F}_k}{\sigma_{k}}\right)^2 , \label{eq:chi2SED} \\ \smallskip
  \textrm{and} & & \chi^2_{i,\mathrm{prof}}  = 
  \sum_{j=1}^{2}\limits
  \sum_{k=1}^{N_{i,j}}\limits \left(\frac{x_i S_{i,j,k} -
      \tilde{S}_{i,j,k}}{\sigma_{i,j,k}}\right)^2 , \label{eq:chi2prof}
\end{eqnarray}
where \mbox{$i = 0$} is for the SED and \mbox{$i \geq 1$} for the profiles, \mbox{$j = 1$} for major axis and \mbox{$j = 2$} for minor axis. In equations
\ref{eq:chi2} to \ref{eq:chi2prof}, the notations used are:

\smallskip
\begin{itemize}
 \item $M$ -- number of images
 \item $N_{0}$ -- number of SED data points
 \item $N_{i,1}$ -- number of profile data points along major axis
 \item $N_{i,2}$ -- number of profile data points along minor axis
 \item $N\dma{tot}$ -- total number of data points, \mbox{$N_0 + \sum_{i=1}^{M}\limits\sum_{j=1}^{2}\limits N_{i,j}$}
 \item $N\dma{free}$ -- number of free parameters
 \item $F_k$, $\tilde{F}_k$ -- modeled and observed SED, respectively
 \item $S_{i,j,k}$, $\tilde{S}_{i,j,k}$ -- modeled and observed surface brightness profiles
 \item $\sigma_k$, $\sigma_{i,j,k}$ -- observed uncertainties
 \item $x_{i}$ -- surface brightness profile scaling factor
 \item $w_i$ -- weight.
\end{itemize}
\smallskip

For the present work, only one profile is used and $w_0$ and $w_1$ are set to 1.0, which means that the profile is given the same weight as the SED.

In order to allow one to find the best-fit model within a given range of parameters of a high dimensional parameter space in reasonable time, one has to compute the value of $\chi^{2}$ for a given set of parameters very fast. Creating the SED and three profiles at different wavelengths for one particular model and comparing them with the observations usually takes \mbox{$\sim 0.1\,\rm{sec}$} on an Intel Xeon E5410 CPU (2.3\,GHz). The exact time depends mostly on the particular set of parameters and the number of wavelengths, at which radial profiles are provided. It hardly depends on the number of data points in the SED and the profiles.

The simulated annealing approach enables us to handle a very large and high dimensional parameter space without sampling the whole range of parameters. Therefore, a random walk on the grid of possible parameters is started at an arbitrary position. The probability $p$ to go a certain step depends on $\chi^{2}_{1}$ of the actual and $\chi^{2}_{0}$ of the the previous step. It is 1 for \mbox{$\chi^{2}_{0} \ge \chi^{2}_{1}$} (fit becomes better) and follows a Boltzmann distribution for \mbox{$\chi^{2}_{0} < \chi^{2}_{1}$} (fit becomes worse):
\begin{eqnarray}
  \label{prob}
  p = \exp\left[-\frac{\chi^{2}_{1} - \chi^{2}_{0}}{T}\right]
\end{eqnarray}
In this equation, the quantity $T$ is a parameter comparable to the Temperature of an ``annealing'' physical system and controls the ability to reach areas in the parameter space that give worse fits. It is chosen at the beginning of each run to be a large value based on experiences from previous runs and is lowered with each successful step of the random walk. The width of each step is chosen from a probability distribution that prefers short steps, while the maximum step width is $1/2$ of the whole range of values for each parameter. The probability to go long steps is lowered with each successful step. With this approach the code is able to reach each position in the parameter space, but prefers regions with low $\chi^{2}$. The run stops, when a position in the parameter space with the following properties is reached:
\begin{itemize}
 \item With the actual value of $T$ the code was not able to leave the position after computing 500 further models.
 \item The actual $\chi^{2}$ is lower than or equal to the best reached $\chi^{2}$ during the whole run.
 \item The actual $\chi^{2}$ is lower than or equal to a maximum $\chi^{2}$ based on experiences from previous runs (input parameter).
\end{itemize}
A set of model parameters that satisfies all of these conditions is considered as a likely global fit. To increase confidence in this result, one can start several runs at different regions of the parameter space (4 in the recent case for the final fitting). Only if all runs reach the same region in the parameter space, we consider the result to be really a global best-fit. If only the first and not more than one other of these stopping criteria is satisfied, the value of $T$ is increased and the code goes on trying to leave the actual position in the parameter space.

Error estimates are done in the context of the simulated annealing approach by starting a new random walk at the best-fit position with a fixed value of $T$ (usually 10\% of the best-fit $\chi^{2}$ in our case). The code counts how often each value of a parameter is reached. This way, the code samples the projection of the probability distribution of the random walk (Eq.~\ref{prob}) on the axis of each parameter. With the knowledge of the value of $T$ one can then compute the probability distribution
\begin{eqnarray}
  p' = \exp\left[-\left(\chi^{2} - \chi^{2}_{\rm best}\right)\right]
\end{eqnarray}
and corresponding levels of confidence.

As a feature of the simulated annealing approach the time it takes to find a best-fit (and uncertainties) does not vary very much with the number of free parameters or the total size of the parameter space to be searched, but depends on the shape of the $\chi^2$~distribution in the parameter space. Weak variations of the $\chi^2$ over a broad range of values of the parameters as well as very degenerate fitting problems require more time, i.e. more models to be computed. For the recent work (SED fitting only) we find the best-fit in a parameter space composed of $\sim 10^{10}$ to $10^{11}$ knots typically in less than 24\,hours including the estimates of the uncertainties. Therefore the code computes about 100.000 to 1.000.000 models and accepts (goes) about 10.000 and 100.000 steps for the fitting and computes $\sim$1.000.000 models and accepts $\sim$300.000 steps for the error estimates.

\end{appendix}

\clearpage
\onecolumn
\begin{landscape}
\setcounter{table}{4}
\begin{table*}
\caption{Results from the fitting based on simulated thermal annealing (for details see Sect.~\ref{fitting})}
\label{results_sand}
\begin{center}
\begin{tabular*}{1.0\linewidth}{c@{\extracolsep{\fill}} cccccccc}
\hline\hline                 
   Parameter & \multicolumn{8}{c}{Best-fit value [$3\sigma$ confidence levels]} \\
\hline \\
\smallskip
   & \multicolumn{8}{c}{HIP\,103389} \\
\hline
   & \multicolumn{2}{c}{Approach~1} & \multicolumn{2}{c}{Approach~2} & \multicolumn{2}{c}{Approach~3} & \multicolumn{2}{c}{Approach~4} \\
  \cline{2-3} \cline{4-5} \cline{6-7} \cline{8-9}
   & Silicate & Mixture & Silicate & Mixture & Silicate & Mixture & Silicate & Mixture \\
\hline
  $R_{\rm{in}}$ [AU]            & 20.7 [8.5 -- 32.2]  & 20.8 [9.1 -- 35.9]   & 18.2 [7.9 -- 24.5]   & 20.9 [8.0 -- 26.6]   & 11.9 [4.2 -- 16.0]  & 13.6 [4.4 -- 19.1]         & 42.3 [12.9 -- 63.3]  & 22.5 [8.6 -- 43.0]   \\
  $R_{\rm{out}}$ [AU]           & 20.7 [15.3 -- 53.1] & 20.9 [15.4 -- 123.3] & 20.0 [16.3 -- 65.1]  & 20.9 [17.3 -- 77.3]  & 12.0 [8.8 -- 37.4]  & 13.6 [10.2 -- 38.5]        & 46.0 [21.3 -- 138.5] & 22.5 [17.9 -- 110.9] \\
  $\alpha$                      & 0.1 [$-2.0$ -- 3.5] & 2.6 [$-2.0$ -- 3.5]  & 0.0 (fixed)          & 0.0 (fixed)          & 0.0 (fixed)         & 0.0 (fixed)                & 0.0 (fixed)          & 0.0 (fixed)          \\
  $a_{\rm{min}}$ [$\um$]        & 6.6 [4.7 -- 8.6]    & 9.2 [4.4 -- 11.4]    & 9.5 [7.8 -- 10.4]    & 12.8 [10.7 -- 13.6]  & 6.1 [2.8 -- 9.8]    & 6.6 [2.6 -- 12.0]          & 4.2 [3.1 -- 7.8]     & 9.3 [3.7 -- 14.9]    \\
  $a_{\rm{max}}$ [$\um$]        & 1000.0 (fixed)      & 1000.0 (fixed)       & 1000.0 (fixed)       & 1000.0 (fixed)       & 1000.0 (fixed)      & 1000.0 (fixed)             & 14.3 [12.7 -- 18.1]  & 22.7 [14.3 -- 28.5]  \\
  $\gamma$                      & 5.0 [4.6 -- 5.0]    & 5.0 [4.5 -- 5.0]     & 7.4 [6.3 -- 10.0]    & 9.0 [6.5 -- 10.0]    & 3.5 (fixed)         & 3.5 (fixed)                & 3.5 (fixed)          & 3.5 (fixed)          \\
  $M_{\rm{dust}}$ [$M_{\oplus}$] & 1.98e-5            & 1.67e-5              & 1.30e-5              & 1.29e-5              & 4.69e-5             & 4.03e-5                    & 4.86e-5              & 1.45e-5              \\
  reduced $\chi^2$              & 4.112               & 3.664                & 0.816                & 0.776                & 10.277              & 8.488                      & 0.628                & 0.760                \\
\hline \\
\smallskip
   & \multicolumn{8}{c}{HIP\,107350} \\
\hline
   & \multicolumn{2}{c}{Approach~1} & \multicolumn{2}{c}{Approach~2} & \multicolumn{2}{c}{Approach~3} & \multicolumn{2}{c}{Approach~4} \\
  \cline{2-3} \cline{4-5} \cline{6-7} \cline{8-9}
   & Silicate & Mixture & Silicate & Mixture & Silicate & Mixture & Silicate & Mixture \\
\hline
  $R_{\rm{in}}$ [AU]            & 15.3 [3.8 -- 28.5]  & 21.5 [7.4 -- 93.0]   & 29.1 [7.9 -- 47.2]   & 30.6 [5.6 -- 44.1]   & 9.6 [3.0 -- 15.6]   & 10.9 [3.0 -- 16.5]         & 37.1 [4.3 -- 54.4]   & 35.2 [7.2 -- 54.2]   \\
  $R_{\rm{out}}$ [AU]           & 15.9 [9.3 -- 227.3] & 21.8 [10.3 -- 276.6] & 31.3 [13.9 -- 113.9] & 32.3 [16.0 -- 138.5] & 9.6 [5.8 -- 33.0]   & 11.0 [9.1 -- 35.7]         & 37.4 [19.2 -- 187.5] & 35.2 [17.4 -- 145.3] \\
  $\alpha$                      &$-0.9$ [$-2.0$ -- 3.5]& 1.4 [$-2.0$ -- 3.5] & 0.0 (fixed)          & 0.0 (fixed)          & 0.0 (fixed)         & 0.0 (fixed)                & 0.0 (fixed)          & 0.0 (fixed)          \\
  $a_{\rm{min}}$ [$\um$]        & 5.9 [4.7 -- 9.0]    & 5.9 [4.3 -- 9.1]     & 6.9 [2.7 -- 10.9]    & 8.2 [3.6 -- 10.7]    & 5.7 [4.8 -- 11.0]   & 5.8 [4.9 -- 11.7]          & 7.8 [1.6 -- 10.4]    & 9.6 [2.2 -- 13.5]    \\
  $a_{\rm{max}}$ [$\um$]        & 1000.0 (fixed)      & 1000.0 (fixed)       & 1000.0 (fixed)       & 1000.0 (fixed)       & 1000.0 (fixed)      & 1000.0 (fixed)             & 7.8 [6.3 -- 13.3]    & 9.6 [6.4 -- 17.4]    \\
  $\gamma$                      & 5.0 [4.0 -- 5.0]    & 5.0 [4.0 -- 5.0]     & 10.0 [6.0 -- 10.0]   & 10.0 [5.9 -- 10.0]   & 3.5 (fixed)         & 3.5 (fixed)                & 3.5 (fixed)          & 3.5 (fixed)          \\
  $M_{\rm{dust}}$ [$M_{\oplus}$] & 4.59e-6            & 6.03e-6              & 1.11e-5              & 9.29e-6              & 1.05e-5             & 9.09e-6                    & 1.54e-5              & 1.08e-5             \\
  reduced $\chi^2$              & 7.048               & 5.672                & 1.652                & 1.568                & 4.229               & 3.869                      & 1.528                & 1.488                \\
\hline \\
\smallskip
   & \multicolumn{8}{c}{HIP\,114948} \\
\hline
   & \multicolumn{2}{c}{Approach~1} & \multicolumn{2}{c}{Approach~2} & \multicolumn{2}{c}{Approach~3} & \multicolumn{2}{c}{Approach~4} \\
  \cline{2-3} \cline{4-5} \cline{6-7} \cline{8-9}
   & Silicate & Mixture & Silicate & Mixture & Silicate & Mixture & Silicate & Mixture \\
\hline
  $R_{\rm{in}}$ [AU]            & 12.8 [6.8 -- 13.9]  & 13.6 [6.8 -- 14.5]  & 12.8 [7.1 -- 13.7]   & 13.5 [7.8 -- 14.9]   & 12.7 [5.5 -- 16.6]  & 14.0 [7.9 -- 19.1]         & 32.5 [9.0 -- 40.1]  & 13.3 [8.4 -- 14.5]  \\
  $R_{\rm{out}}$ [AU]           & 12.8 [12.1 -- 39.8] & 13.8 [12.9 -- 34.7] & 12.8 [12.1 -- 23.2]  & 13.8 [12.9 -- 24.9]  & 12.7 [10.5 -- 25.6] & 14.1 [11.0 -- 30.8]        & 34.5 [26.9 -- 81.4]  & 13.8 [12.9 -- 22.9] \\
  $\alpha$                      & 2.9 [$-2.0$ -- 3.5] & 0.0 [$-2.0$ -- 3.5] & 0.0 (fixed)          & 0.0 (fixed)          & 0.0 (fixed)         & 0.0 (fixed)                & 0.0 (fixed)          & 0.0 (fixed)         \\
  $a_{\rm{min}}$ [$\um$]        & 9.6 [8.9 -- 10.2]   & 12.9 [11.9 -- 13.8] & 9.6 [8.9 -- 10.2]    & 12.9 [11.9 -- 13.8]  & 6.4  [3.6 -- 9.9]   & 7.4 [3.0 -- 11.7]          & 3.2 [2.6 -- 4.6]     & 11.8 [10.7 -- 13.1] \\
  $a_{\rm{max}}$ [$\um$]        & 1000.0 (fixed)      & 1000.0 (fixed)      & 1000.0 (fixed)       & 1000.0 (fixed)       & 1000.0 (fixed)      & 1000.0 (fixed)             & 24.2 [10.0 -- 27.3]  & 43.8 [10.0 -- 53.2] \\
  $\gamma$                      & 4.7 [4.5 -- 5.0]    & 4.7 [4.4 -- 5.0]    & 4.7 [ 4.4 -- 5.2]    & 4.7 [4.4 -- 5.1]     & 3.5 (fixed)         & 3.5 (fixed)                & 3.5 (fixed)          & 3.5 (fixed)         \\
  $M_{\rm{dust}}$ [$M_{\oplus}$] & 2.19e-5            & 2.16e-5             & 2.19e-5              & 2.16e-5              & 8.82e-5             & 7.36e-5                    & 5.93e-5              & 1.61e-5             \\
  reduced $\chi^2$              & 0.568               & 0.464               & 0.284                & 0.232                & 3.291               & 2.333                      & 0.132                & 0.232               \\
\hline
\end{tabular*}
\end{center}
\end{table*}
\end{landscape}
\twocolumn

\end{document}